\documentclass[11pt]{article}
\usepackage[T1]{fontenc}
\usepackage[utf8]{inputenc}
\usepackage{lmodern,amsmath,amssymb,amsthm,mathtools}
\usepackage[margin=1in]{geometry}
\usepackage{enumitem,booktabs,microtype}
\usepackage{xcolor}
\definecolor{newblue}{rgb}{0.19,0.55,0.91}
\usepackage[colorlinks=true,linkcolor=blue,citecolor=newblue,urlcolor=blue]{hyperref}
\setlist{itemsep=3pt,topsep=5pt}

\newtheorem{theorem}{Theorem}[section]
\newtheorem{lemma}[theorem]{Lemma}
\newtheorem{proposition}[theorem]{Proposition}

\theoremstyle{definition}
\newtheorem{definition}[theorem]{Definition}
\theoremstyle{remark}

\newcommand{\eps}{\varepsilon}
\newcommand{\E}{\mathbb E}
\newcommand{\Pp}{\mathbb P}
\newcommand{\Var}{\operatorname{Var}}
\newcommand{\ind}{\mathbf 1}
\newcommand{\med}{\operatorname{median}}
\newcommand{\sgn}{\operatorname{sgn}}
\newcommand{\str}{\operatorname{str}}
\newcommand{\norm}[1]{\lVert #1\rVert}
\newcommand{\Sel}{\mathsf S}
\newcommand{\Diag}{\mathsf D}
\newcommand{\Est}{\mathsf T}
\newcommand{\id}{\operatorname{id}}
\newcommand{\bit}{\operatorname{bit}}

\title{Near-Optimal Space and Time Cut Sparsification \\ in Turnstile Streams}

\author{
Honghao Lin\footnote{Google Research}
\and
Vahab Mirrokni\footnote{Google Research}
\and
David P. Woodruff\footnote{Google Research and Carnegie Mellon University}
}

\begin{document}
\maketitle

\begin{abstract}
A seminal result of Ahn, Guha, and McGregor (PODS'12) showed that a cut
sparsifier of an unweighted graph can be computed from a near-linear
number of linear measurements. Later work extended this to spectral
sparsifiers (Kapralov et al., FOCS'14) and to near-linear recovery time
(Kapralov et al., SODA'20). For weighted graphs, these sketches rely on
grouping edges by weight, which general turnstile streams do not allow.
Chen, Khanna, and Li (FOCS'22) removed this limitation, using
$\widetilde O(n/\eps^3)$ bits of space.

We improve the space to $\widetilde O(n/\eps^2)$ bits. Our algorithm
makes one pass over a general turnstile stream of a weighted graph,
processes each update in $\widetilde O(1)$ time, and recovers, in
$\widetilde O(n/\eps^2)$ time, a reweighted subgraph with
$O(n\eps^{-2}\log^2n)$ edges that preserves every cut within a factor of
$1\pm\eps$. Its space, update time, and recovery time are all optimal up
to polylogarithmic factors. 
\end{abstract}

\section{Introduction}\label{sec:intro}

Cut sparsification, introduced by Bencz\'ur and Karger~\cite{BK96}, has
been profoundly influential. Given a weighted graph $G=(V,E,w)$ with
$n=|V|$ vertices, nonnegative edge weights $w_e$, and an error parameter
$\eps>0$, a $(1\pm\eps)$-cut sparsifier of $G$ is a reweighted subgraph
$H$ on the same vertex set that approximates the weight of every cut of
$G$ within a factor of $1\pm\eps$. They proved that every undirected
graph admits such a sparsifier with only
$O(n\log n/\eps^2)$ edges. This breakthrough was later extended to the
stronger notion of spectral sparsifiers~\cite{ST11}, for which the edge
bound was improved to $O(n/\eps^2)$~\cite{BSS12}. A rich line of
follow-up work has given alternative constructions and refined bounds
for both cut and spectral
sparsifiers~\cite{FHHP11,KP12,ST04,SS11,LeeS17,CKST19}. Beyond its
theoretical interest, sparsification has become a basic building block
of modern graph algorithms, including algorithms for maximum flow and
minimum cut~\cite{BK15,She13,KLOS14,Peng16}, solvers for
graph-structured linear systems~\cite{ST14,CKM+14,JS25}, and methods for
graph clustering~\cite{CSWZ16}.

Sparsifiers also serve as compact synopses of a graph, a viewpoint that
is central to graph sketching and data
streaming~\cite{AGM12,mcgregor2014graph}. A cut query asks for the total
weight of the edges crossing a partition of the vertex set, a basic
aggregate query over the edge set. A single cut sparsifier answers all
such queries simultaneously up to a multiplicative error, using space
nearly proportional to the number of vertices rather than the number of
edges~\cite{BK96,BK15}. Downstream procedures for partitioning,
connectivity, minimum cut, and flow can then work with this summary
without materializing the full graph.

This perspective makes the streaming and linear-sketching models
particularly natural. A streaming algorithm processes the edge set as a
sequence of updates and maintains only a small summary of the graph. A
linear sketch is a summary that is a linear function of the edge-weight
vector, so it handles insertions and deletions alike. A seminal result
of Ahn, Guha, and McGregor~\cite{AGM12} showed that a cut sparsifier of
an unweighted undirected graph can be computed from only
$\widetilde O(n/\eps^2)$ linear measurements. This line of work was
subsequently extended to spectral sparsifiers~\cite{KLM+17} and to
faster recovery algorithms~\cite{KMM+20}. On the lower-bound side,
Andoni et al.~\cite{ACK+15} proved that, for $\eps>1/\sqrt n$, any data
structure that preserves all cuts simultaneously needs
$\Omega(n/\eps^2)$ bits, and Carlson et al.~\cite{CKST19} strengthened
this to $\Omega(n\log n/\eps^2)$ bits for $\eps=\omega(n^{-1/4})$.

The general turnstile model captures a more flexible form of dynamic
graph data. Edge weights may represent aggregate quantities such as
counts, frequencies, capacities, or affinities, and updates may both
increase and decrease them. Such updates arise naturally when records
are inserted or deleted, when late-arriving records correct earlier
aggregates, or when transactions are rolled back. Prior linear sketches
for weighted graphs handle a more restrictive dynamic model by grouping
edges into geometric weight classes~\cite{AGM12,KLM+17,KMM+20}. This
requires every deletion of an edge to reveal the weight used at its
insertion, so that the edge stays in the same weight class throughout
the stream.

Grouping edges by weight, however, is inherently nonlinear, so it cannot
be implemented in the general turnstile model, where the input is an
arbitrary sequence of edge-weight updates. To overcome this limitation,
Chen, Khanna, and Li~\cite{CKL22} gave a linear sketch for weighted
undirected graphs with $\widetilde O(n/\eps^3)$ measurements, yielding a
turnstile streaming algorithm with the same space bound. The dependence
on $\eps$ is central here: the offline sparsification
bounds~\cite{BK96,BK15,BSS12} and the sketching lower
bounds~\cite{ACK+15,CKST19} suggest that $\widetilde O(n/\eps^2)$ space
is the right target. Improving this dependence matters most in
high-accuracy regimes, where $\eps$ is small and the extra $1/\eps$
factor dominates the space. Time is the other resource. Ideally, each
update takes polylogarithmic time and the sparsifier is recovered in
time nearly linear in its size. For dynamic streams with weight
grouping, Kapralov et al.~\cite{KMM+20} achieved such a recovery time.
This leads to the following question:
\begin{quote}
\emph{Can we design a cut sparsifier algorithm for general turnstile
streams that matches the near-optimal $\widetilde O(n/\eps^2)$ space
bound of the offline setting and runs in near-linear time?}
\end{quote}

\subsection{Our result}\label{sec:result}

We answer this question in the affirmative. The input is a weighted
graph $G=(V,E,w)$ on $n$ vertices, given by a turnstile stream of $M$
integer edge-weight updates of magnitude at most $U$, with final weights
in $\{0,1,\ldots,U\}$. Running times refer to the word RAM with
$\Theta(\log n)$-bit words, and Section~\ref{sec:prelims} gives the
precise model.

\begin{theorem}\label{thm:main}
Let $0<\eps<1$ and $0<\delta<1/4$, and assume
$M,U,\delta^{-1}\le n^{O(1)}$. There is a one-pass linear-sketch
algorithm using $\widetilde O(n/\eps^2)$ bits of space,
$\widetilde O(1)$ worst-case time per update, and
$\widetilde O(n/\eps^2)$ worst-case recovery time that, with probability
at least $1-\delta$, returns a positive reweighting $H$ of
$O(n\eps^{-2}\log^2n)$ edges of $G$ such that
\[
 (1-\eps)\,w(S,V\setminus S)\le w_H(S,V\setminus S)
       \le(1+\eps)\,w(S,V\setminus S)\qquad\text{for every }S\subseteq V.
\]
\end{theorem}

The recovery time is the time from the end of the stream until $H$ is
returned, so the total running time is $\widetilde O(M+n/\eps^2)$.
We have formally verified
Theorem~\ref{thm:main} in Lean\footnote{The formalization takes the approximate Gomory--Hu tree
algorithm of~\cite{LNPS23} (Theorem~\ref{thm:approx-gh}) as a
hypothesis. All other results used in the paper are proved within the
formalization.}, and the formalization is available
\href{https://drive.google.com/file/d/1DKUPH59AY8SsRgvutVgO8HemCJ7uOCW3/view?usp=sharing}{here}.
Theorem~\ref{thm:main} improves the space bound of Chen, Khanna, and
Li~\cite{CKL22} by a factor of $1/\eps$; Table~\ref{tab:comparison}
summarizes the comparison.

The space, update time, and recovery time of our algorithm are all
optimal up to polylogarithmic factors. This is clear for the update
time. For $\eps>1/\sqrt n$, Andoni et al.~\cite[Theorems~2.3
and~2.4]{ACK+15} showed that any data structure that approximately
preserves all cuts needs $\Omega(n/\eps^2)$ bits, and that some
$n$-vertex graphs require $\Omega(n/\eps^2)$ edges in every
$(1\pm\eps)$-cut sparsifier, so writing the output alone takes
$\Omega(n/\eps^2)$ time. Carlson et al.~\cite{CKST19} strengthened the
space lower bound to $\Omega(n\log n/\eps^2)$ bits for
$\eps=\omega(n^{-1/4})$. For smaller $\eps$, storing the graph exactly
is cheaper, and our algorithm does so whenever $n/\eps^2\ge\binom n2$
(Appendix~\ref{app:precision}). Its space and recovery time are
therefore $\widetilde O(\min\{n^2,n/\eps^2\})$ for every $\eps$, which
is again optimal: for $n\ge5$ and $\eps\le1/\sqrt n$, the lower bounds
above, applied with accuracy $2/\sqrt n$, give $\Omega(n^2)$.

\begin{table}[ht]
\centering
\begin{tabular}{@{}lllll@{}}
\toprule
Reference & Input model & Space (bits) & Update time & Recovery time\\
\midrule
\cite{AGM12,KLM+17} & dynamic, weight grouping
  & $\widetilde O(n/\eps^2)$ & --- & $\operatorname{poly}(n)$\\
\cite{KMM+20} & dynamic, weight grouping
  & $\widetilde O(n/\eps^2)$ & $\widetilde O(1/\eps^2)$
  & $\widetilde O(n/\eps^2)$\\
\cite{CKL22} & general turnstile & $\widetilde O(n/\eps^3)$ & ---
  & $\operatorname{poly}(n)$\\
This work & general turnstile & $\widetilde O(n/\eps^2)$
  & $\widetilde O(1)$ & $\widetilde O(n/\eps^2)$\\
\midrule
Lower bounds & & $\Omega(n\log n/\eps^2)$ & & $\Omega(n/\eps^2)$\\
\bottomrule
\end{tabular}
\caption{Bounds for $(1\pm\eps)$-cut sparsification of weighted
$n$-vertex graphs. A dash marks an update time that the cited works do
not state. The space lower bound $\Omega(n\log n/\eps^2)$ holds for
$\eps=\omega(n^{-1/4})$~\cite{CKST19}; for $\eps>1/\sqrt n$, the weaker
bound $\Omega(n/\eps^2)$ holds~\cite[Theorem~2.3]{ACK+15}. The
recovery-time lower bound is the output size~\cite[Theorem~2.4]{ACK+15}
and holds for $\eps>1/\sqrt n$.}
\label{tab:comparison}
\end{table}

\paragraph{Comparison with for-each cut sketching.}
Liu, Nguyen, Shah, and Woodruff~\cite{LNSW26mincut} recently gave an
$\widetilde O(n/\eps)$-bit for-each cut sketch in general turnstile
streams, which yields $(1+\eps)$-approximate global minimum cuts and
fixed-$k$ partitions in $\widetilde O(n/\eps)$ space. Their data
structure estimates the weight of a single fixed cut query and cannot
output a sparse graph or preserve all cuts simultaneously, which require
$\Omega(n\log n/\eps^2)$ bits by the lower bounds above. It also cannot
solve $(1+\eps)$-approximate $s$-$t$ minimum cut or Gomory--Hu tree.
As they point out, the optimal space for preserving all cuts in general
turnstile streams was not known, since the best bound,
$\widetilde O(n/\eps^3)$~\cite{CKL22}, exceeds the lower bounds by a
factor of $1/\eps$. Theorem~\ref{thm:main} closes this gap.
Moreover, their minimum-cut and $k$-partition algorithms use a
constant-accuracy for-all cut sparsifier for general turnstile streams
as a black box, for which they cite~\cite{CKL22}. As a result, their
update time is $t$ plus a polylogarithmic term, where $t$ is the update
time of this sparsifier, and the sparsifier takes polynomial time to
recover. With Theorem~\ref{thm:main} at constant $\eps$, their
algorithms run in $\widetilde O(1)$ worst-case time per update, and the
sparsifier is recovered in $\widetilde O(n)$ time.

\paragraph{Presentation.}
For ease of exposition ,Sections~\ref{sec:algorithm}--\ref{sec:allcuts} present a
simpler version of the algorithm, which runs in polynomial time and uses
idealized randomness. Section~\ref{sec:nearlinear} then implements every
step in near-linear time and generates all randomness from short seeds,
which gives Theorem~\ref{thm:main}.

\subsection{Technical overview}\label{sec:overview}

We now describe the main ideas behind Theorem~\ref{thm:main}. For a
vertex pair $e=uv$, let $\lambda_e$ denote its connectivity, the minimum
weight of a cut separating $u$ and $v$.

\paragraph{Why the previous approach loses a factor of $1/\eps$.}
The starting point is the linear sketching algorithm of~\cite{CKL22},
which implements the Bencz\'ur--Karger strength-based sampling
scheme~\cite{BK96,BK15} in a general turnstile stream. In this scheme,
each edge $e$ is sampled with probability
$p_e=\min\{1,\rho w_e/\str(e)\}$ and reweighted by $1/p_e$, where
$\rho=\Theta(\log n/\eps^2)$. The strength $\str(e)$ is the largest $k$
such that both endpoints of $e$ lie in a vertex-induced subgraph whose
minimum cut is at least $k$. Intuitively, edges with smaller strength
lie on sparser cuts and therefore need to be sampled with higher
probability. Since $\sum_ew_e/\str(e)\le n-1$, this sampling rule
produces only $\widetilde O(n/\eps^2)$ sampled edges.

The challenge is that in a turnstile stream, after an edge is sampled,
we also need to know its final weight $w_e$ and its inverse sampling
probability $1/p_e$, which together give its output weight $w_e/p_e$.
The algorithm of~\cite{CKL22}
recovers sampled edges using $\ell_1$ samplers. However, to output a
sparsifier with $(1\pm\eps)$ accuracy, their analysis estimates each
sampled edge weight to $(1\pm\eps)$ multiplicative accuracy. This
accuracy requirement costs an additional $1/\eps$ factor in the $\ell_1$
sampler. Since the sampling rate already contributes the factor
$\rho=\widetilde O(1/\eps^2)$, the total space becomes
$\widetilde O(n/\eps^3)$.

\paragraph{Constant-accuracy estimates suffice.}
Our goal is to avoid estimating each sampled weight to relative error
$\eps$. Intuitively, if the recovered weights are unbiased, their
errors should average out across a cut, just as the sampling errors do.
We use connectivity-based sampling (e.g.,~\cite{AGM12,FHHP11}): each
edge $e$ is sampled with probability
$p_e=\min\{1,\beta w_e/\lambda_e\}$ and reweighted by $1/p_e$, where
$\beta=\widetilde\Theta(\eps^{-2})$ is a sampling budget. Since
$\sum_ew_e/\lambda_e\le n-1$, this samples $O(n\beta)$ edges in
expectation. If $p_e<1$, a sampled edge receives weight
$w_e/p_e=\lambda_e/\beta$. The sampling error on a cut is therefore a
sum of independent zero-mean terms, one per edge of the cut, each of
magnitude at most $\lambda_e/\beta$ when $p_e<1$. For a cut of weight $W$,
every edge $e$ of the cut has $\lambda_e\le W$, so these terms are at
most $W/\beta$ and their variances sum to at most $W^2/\beta$. With
$\beta=\widetilde\Theta(\eps^{-2})$, Bernstein's inequality keeps the
sampling error within $\pm\eps W$, which is why connectivity sampling
gives a $(1\pm\eps)$-cut sparsifier. Now suppose that the weight of a
sampled edge is estimated with an unbiased error of order $w_e$, that
is, only to constant relative accuracy. After reweighting by $1/p_e$,
this error is also of order $\lambda_e/\beta$, the same scale as the
sampling terms. If these errors behaved like independent zero-mean
terms, they would also sum to $O(\eps W)$, and constant-accuracy
estimates would still yield a $(1\pm\eps)$-cut sparsifier, at the cost of
only a constant factor in $\beta$. This is what we prove. The difficulty
is to show it for all cuts simultaneously: there are exponentially many
cuts, so a direct union bound over them fails. Moreover, the estimation
errors of different edges are not independent.

\paragraph{Implementing connectivity sampling in a turnstile stream.}
The final weights, and hence the sampling probabilities, are unknown
while updates arrive. As is standard for linear
sketches~\cite{AGM12,KLM+17,KMM+20}, we maintain nested subsamples of
all coordinates at rates $1,1/2,1/4,\ldots$, called layers: each
coordinate survives from one layer to the next independently with
probability $1/2$, as determined by its random survival bits. We choose
the layer of each edge after the stream. As
in the improved algorithm of~\cite{AGM12}, we read the sampled edges
through cut queries on vertex sketches: summing the sketches of the
signed incidence vectors over a vertex set $S$ cancels the edges inside
$S$ and leaves a sketch of the sampled edges that cross $S$. The query
cuts come from a Gomory--Hu tree~\cite{GH61} of a constant-accuracy
sparsifier $H_0$, which we call the coarse sparsifier. It is computed
from an independent sketch, either by~\cite{CKL22} in polynomial time or
by our construction in Section~\ref{sec:coarse} in near-linear time. For
every pair $e$, the tree supplies an estimate
$\kappa_e=\Theta(\lambda_e)$ and a query cut $S_e$ that separates the
endpoints of $e$ and has weight $O(\kappa_e)$.

\paragraph{From recovering edges to estimating weights.}
In the unweighted setting of~\cite{AGM12}, the sampled edges crossing a
query cut can be recovered exactly. In a weighted turnstile stream, we
must instead estimate their weights, and these estimates also determine
the layers. At layer $k$, the sampled edges crossing $S_e$ have total
weight $O(2^{-k}\kappa_e)$ in expectation, so a hash table with
$\Theta(\beta)$ buckets reads the sampled weight of $e$ up to an additive
error that is small compared with the scale $t_e(k)=2^{-k}\kappa_e/\beta$.
The decoder takes $e$ at the first layer where this reading exceeds a
constant multiple of $t_e(k)$; for edges with $p_e<1$, this happens when
$2^{-k}$ is of order $\beta w_e/\kappa_e$, as desired.

In offline sampling, each $p_e$ is fixed in advance. Here the reading
contains collision noise from the other sampled edges, so the layer of
$e$ may depend on which other edges survived. When all readings are
accurate, however, this noise only decides between two adjacent layers:
$e$ is too light to pass below a baseline layer $k_e$, which depends
only on $w_e$ and $\kappa_e$, and it passes at layer $k_e+1$ whenever it
survives. Moreover, whether $e$ is taken at layer $k_e+1$ instead of
$k_e$ is determined by the sketch at layer $k_e$, before the survival
coin of $e$ at layer $k_e+1$ is used. The sampling error therefore
splits into a baseline error, a sum of independent terms bounded by
Bernstein's inequality, and an extra-layer error, a martingale whose
increments are fresh fair coins times predetermined coefficients,
bounded by Freedman's inequality.

\paragraph{Separating selection from estimation.}
If the reading that selects $e$ were also used to estimate $w_e$, the
estimate would be biased, because $e$ is selected exactly when its
reading is large. We therefore keep the randomness used for estimation
separate from the randomness used for selection. Each vertex stores
three tables at every layer:
\begin{enumerate}
\item a \emph{selection table}, which chooses the edge and its layer;
\item a \emph{diagnostic table}, which finds a row in which the bucket of
the selected edge has small collision noise; and
\item an \emph{estimation table}, which estimates the selected weight
with independent evaluation signs.
\end{enumerate}
The selection and diagnostic tables do not involve the evaluation
signs. The chosen edges, layers, and rows are therefore independent of
these signs, and every weight estimate remains unbiased given these
choices. However, the estimation errors of
different edges are not independent, as we discuss next.

\paragraph{Estimation errors share random signs.}
Recall the estimation table. At each layer, it has $R$ independent rows;
each row hashes the sampled edges into $\Theta(\beta)$ buckets, and each
bucket keeps $R$ counters, which sum the same edges with independent
evaluation signs. Thus every evaluation sign $\sigma_f^{\pi,l}$ is
indexed by an edge $f$, a layer and row $\pi$, and a repetition
$l\in[R]$. For a selected edge $e$, let $j_e$ be its layer and $\pi_e$
its layer together with the row chosen by the diagnostic; we fix these
choices from now on. The estimate of $e$ multiplies the $R$ counters of
its bucket at $\pi_e$ by its decoding signs $\sigma_e^{\pi_e,l}$ and
averages over $l$, so
\[
 \widehat w_e-w_e=\frac1R\sum_{l=1}^R\sigma_e^{\pi_e,l}
   \sum_{f\ne e}v_e(f)\,\sigma_f^{\pi_e,l},
\]
where the collision vector $v_e$ has an entry $\pm w_f$ for every other
sampled edge $f$ that crosses $S_e$ and lands in the bucket of $e$, and
zero elsewhere. We call the inner sum the collision noise of $e$. The
estimation error on a cut, the sum of $2^{j_e}(\widehat w_e-w_e)$ over
its selected edges, is therefore a quadratic form in the evaluation
signs, and a decoding sign of one edge may appear in the collision noise
of another. The key observation is a separation within each
connectivity class $E_\nu=\{e:K\le\kappa_e<2K\}$, where $K=2^\nu$: for
selected edges $e\ne f$ of the same class, the sign
$\sigma_f^{\pi_e,l}$ with which $f$ enters the collision noise of $e$ is
a decoding sign of $f$ only if $\pi_e=\pi_f$, and in that case
$v_e(f)=0$, because $w_f$ exceeds the bound on $\norm{v_e}_2$ guaranteed
by the diagnostic. Within a class, no decoding sign therefore appears in
any collision noise. Conditioning on all other signs, the contribution
of the class to the error on a cut is a linear form
\[
 \frac1R\sum_e\sum_{l=1}^R c_e^l\,\sigma_e^{\pi_e,l},\qquad
 c_e^l=2^{j_e}\sum_{f\ne e}v_e(f)\,\sigma_f^{\pi_e,l},
\]
in independent decoding signs, where $e$ ranges over the selected edges
of the class that cross the cut, and a single event controls the energy
$\sum_{e,l}(c_e^l)^2$ for all cuts at once. Across classes, a decoding
sign of one class may enter the collision noise of another, so we apply
the conditional bound class by class and combine the classes by a union
bound, which needs no independence between them.

\paragraph{From one cut to all cuts.}
On a fixed cut, the baseline, extra-layer, and estimation errors
therefore all concentrate. To handle all cuts simultaneously, the
standard tool is Karger's cut counting (see~\cite{karger00}
and~\cite[Theorem~2.1]{FHHP11}): at most $n^{2\alpha}$ cuts have weight
at most $\alpha$ times the minimum cut, so a failure probability of
$n^{-\Omega(\alpha)}$ per cut suffices. We apply this idea to each
connectivity class $E_\nu$ separately. The projection of a cut onto
$E_\nu$ is the set of edges of $E_\nu$ that cross the cut. Cuts with the
same projection incur the same error from this class, so we take a union
bound over the distinct projections. For a projection $F$, let $W_F$ be
the minimum weight of a cut with projection $F$. Since every edge of
$E_\nu$ has connectivity $\Omega(K)$, the projection-counting theorem
of~\cite{FHHP11} shows that at most $n^{O(\alpha)}$ projections have
$W_F\le\alpha K$. We can therefore afford failure probability
$\exp\{-\Theta(W_F\log n/K)\}$ for $F$.

\paragraph{A mass-sensitive union bound.}
A projection $F$ comes with two different weights. The weight $W_F$ of
the lightest cut with projection $F$ determines how many such
projections there are, and hence the failure probability we can afford.
The mass $m_F=w(F)$ of the projection itself, which may be much smaller,
determines the variance: each error term on $F$ collects contributions
of size $O(K/\beta)$ from edges of total weight $m_F$, so its variance
is of order $m_FK/\beta$. Hence, except with probability
$\exp\{-\Theta(W_F\log n/K)\}$, the error on $F$ is at most
\[
 O\Bigl(\sqrt{m_FW_F\log n/\beta}+W_F\log n/\beta\Bigr).
\]
Now consider a cut of weight $W$. The masses of its projections onto the
different classes sum to $W$, and each projection $F_\nu$ satisfies
$W_{F_\nu}\le W$, since the cut itself has projection $F_\nu$. Summing
the bounds over the at most $J=O(\log n)$ classes and applying
Cauchy--Schwarz gives a total error of
$O(W\sqrt{J\log n/\beta}+JW\log n/\beta)$, which is at most $\eps W$
for $\beta=\Theta(\eps^{-2}\log^2n)$. This yields the $(1\pm\eps)$
guarantee with $O(n\eps^{-2}\log^2n)$ output edges and
$\widetilde O(n/\eps^2)$ bits of space. Had we bounded the variance by
$W_FK/\beta$, using the cut weight in place of the mass, the total would
be $O(JW\sqrt{\log n/\beta})$, and $\beta$ would need an extra
logarithmic factor.

\paragraph{Near-linear time.}
The algorithm described so far runs in polynomial time: its decoder
tests all $\Theta(n^2)$ vertex pairs, and the coarse sparsifier $H_0$,
the constant-accuracy sparsifier whose Gomory--Hu tree supplies the
query cuts, is obtained from~\cite{CKL22} in polynomial time. Moreover, the
analysis assumes fully independent survival bits and evaluation signs.
Section~\ref{sec:nearlinear} removes these obstacles. Two steps
build on existing tools. First, since the analysis needs
connectivity estimates only within constant factors, exact Gomory--Hu
trees can be replaced by approximate ones~\cite{LNPS23}
(Section~\ref{sec:query-forest}). Second, the pseudorandom generator of Kapralov et
al.~\cite{KMM+20} produces any requested survival bit or evaluation sign
from a short seed in polylogarithmic time, which gives $\widetilde O(1)$
update time (Section~\ref{sec:tapes}). The two main obstacles, however,
require new ideas, which we describe next.

\paragraph{Candidate discovery.}
At most $O(n\beta)$ pairs are selected, but finding them by testing all
$\Theta(n^2)$ pairs is too slow. Instead, we read candidate pairs
directly off the selection buckets (Section~\ref{sec:discovery}). If one
pair dominates a bucket, the bucket has the sign of that pair's
contribution. Next to each selection counter, we therefore store
$d=\lceil\log_2N\rceil$ bit counters, where $N=\binom n2$ is the number
of vertex pairs, and the $j$th bit counter
multiplies the contribution of each pair by $\pm1$ according to the
$j$th bit of its identifier. Comparing the signs of the bit counters
with the sign of the base counter reveals the identifier of the
dominating pair bit by bit. This encoding is standard in sparse
recovery; to keep the collision noise small in all $d+1$ counters of a
bucket, we widen the selection table by a logarithmic factor. Scanning
the buckets of the at most $n-1$ query cuts yields
$\widetilde O(n/\eps^2)$ candidates.

The key property is that the candidates include every selected edge. A
selected edge $e$ is taken at layer $k_e$ or $k_e+1$, so it survives its
baseline layer $k_e$, where its weight is at least a constant fraction
of the scale $t_e(k_e)$. With high probability, some row has small
collision noise in the bucket of $e$ at its query cut $S_e$, so $e$
dominates that bucket and is decoded there. Every candidate then goes
through exactly the same selection test, diagnostic, and estimate as
before, which also rejects candidates that pair enumeration would not
select. Candidate discovery
therefore returns exactly the output of pair enumeration. The analysis
applies unchanged, while recovery from the tables takes
$\widetilde O(n/\eps^2)$ time instead of $\Theta(n^2)$ selection
queries.

\paragraph{A near-linear-time coarse sparsifier.}
Our algorithm needs, for every pair $e$, a query cut $S_e$ that
separates the endpoints of $e$ and whose weight is within a constant
factor of $\lambda_e$, and it needs all of them in near-linear time.
Such query cuts can be read off an approximate Gomory--Hu tree of any
constant-accuracy sparsifier of $G$. The sketch of~\cite{CKL22} gives
such a sparsifier, but its recovery takes polynomial time. One could
instead run our own algorithm with constant accuracy, since its recovery
takes near-linear time. However, this is circular: our recovery
procedure itself needs query cuts.

The key fact that breaks the circularity is that query cuts need to be
accurate only within constant factors. If every cut of a graph $X'$ is
within a constant factor of the same cut of $X$, then so is every
connectivity, and accurate query cuts for $X$ remain accurate for $X'$. We
therefore follow a path of graphs that starts at a graph whose query
cuts are explicit, ends essentially at $G$, and changes every cut by only
a constant factor per step. For each graph on the path, we run our own
recovery procedure at constant accuracy, with query cuts computed from
the sparsifier recovered for the previous graph
(Section~\ref{sec:coarse}).

Concretely, write $w\in\{0,\ldots,U\}^N$ for the weight vector of $G$,
and let $w_{\rm star}\in\{0,1\}^N$ be the spanning star with center $1$,
whose edges $\{1,u\}$ have weight one. Let $A$ be a large scaling
factor, and let $m_0>m_1>\cdots>m_L$ with
$m_{i+1}\ge64m_i/65$. The graphs
\[
 W_i=Aw+m_iw_{\rm star}\qquad(0\le i\le L)
\]
form a \emph{regularization path}: the star acts as a regularizer that
dominates $W_0$ and is negligible in $W_L$. Consecutive graphs differ by
at most a factor $65/64$ in every coordinate, and hence in every cut. To
lower the multiple from $m_0=\operatorname{poly}(nU)$ to $m_L=O(1)$, the
path therefore takes $L=O(\log(nU))$ steps. We recover constant-accuracy
sparsifiers $Y_0,\ldots,Y_L$ in turn; each $Y_i$ is the output of our
recovery procedure, run at accuracy $1/64$ on a sketch of $W_i$ with
accurate query cuts for $W_i$.
\begin{itemize}
\item \emph{Start.} Let $S$ be a vertex set that does not contain the
center. The star contributes $m_0|S|$ to the cut of $S$ in $W_0$, while
$Aw$ contributes less than $A|S|nU$, which is at most $m_0|S|/128$ for
$m_0=128nAU$. Thus every cut of $W_0$ is within a factor $1+1/128$ of its
star part. The minimum cuts of the star are explicit: the cheapest way
to separate two vertices is to cut off a single leaf. These cuts are
therefore accurate query cuts for $W_0$, and we use them to recover $Y_0$.
\item \emph{Step.} The graph $W_{i+1}$ is available only through its
sketch, but $Y_i$ is an explicit graph with $\widetilde O(n)$ edges, in
which we can compute near-minimum cuts. Every cut of $Y_i$ is within a
constant factor of the same cut of $W_{i+1}$, since $Y_i$ approximates
$W_i$ and $W_i$ differs from $W_{i+1}$ by at most a factor $65/64$.
Hence a near-minimum cut separating two vertices in $Y_i$ is also a
near-minimum cut separating them in $W_{i+1}$, that is, an accurate query
cut. We compute these cuts from an approximate Gomory--Hu tree of $Y_i$,
and with them we run our recovery procedure on the sketch of $W_{i+1}$
to obtain $Y_{i+1}$. Errors
do not accumulate along the path: $Y_{i+1}$ has accuracy $1/64$ with
respect to $W_{i+1}$, and $Y_i$ only tells the recovery where to look.
\item \emph{End.} The factor $A$ makes every cut of positive weight in
$Aw$ at least $A$, so the remaining star, of weight $m_L=O(1)$, is
negligible on such cuts. Discarding the edges of $Y_L$ that join
different components of $G$ and dividing by $A$ gives a constant-factor
approximation of $w$. Its query cuts are accurate for $w$ itself, and a
final run of our recovery procedure on a sketch of $w$ returns a
sparsifier supported on edges of $G$.
\end{itemize}
All these recoveries take place after the stream. Each needs a sketch
whose randomness is independent of its query cuts, so we maintain $L+2$
independent copies of the three-table sketch of $w$ at constant
accuracy, each with $\widetilde O(n)$ bits, in the same single pass. By
linearity, the sketch of $W_i$ is $A$ times the sketch of $w$ plus
$m_i$ times the sketch of $w_{\rm star}$, which we add after the
stream. This proves Theorem~\ref{thm:coarse}: a one-pass sketch with
$\widetilde O(n)$ bits of space, $\widetilde O(1)$ update time, and
$\widetilde O(n)$ recovery time that returns a constant-accuracy
sparsifier. The construction is self-contained: it uses only the
recovery procedure of this paper and the tools above, and in particular
does not rely on~\cite{CKL22}.

\subsection{Related work}\label{sec:related}

\paragraph{Offline sparsification.}
Cut sparsifiers were introduced by Bencz\'ur and
Karger~\cite{BK96,BK15}, who sample each edge with probability inversely
proportional to its strength. Fung, Hariharan, Harvey, and
Panigrahi~\cite{FHHP11} gave a general framework showing that sampling
by connectivity, among other measures, also yields cut sparsifiers; our
sampling rule and our projection counting both come from their work.
Spectral sparsification was introduced by Spielman and
Teng~\cite{ST04,ST11}. Spielman and Srivastava~\cite{SS11} sampled edges
by effective resistances, and Batson, Spielman, and
Srivastava~\cite{BSS12} obtained sparsifiers with $O(n/\eps^2)$ edges,
later constructed in nearly-linear time~\cite{LeeS17}. Further
constructions and related sketches include~\cite{KP12,JambulapatiS18,ChuGPSSW23}.

\paragraph{Linear sketching and dynamic streams.}
Ahn, Guha, and McGregor~\cite{AGM12} gave linear sketches for cut
sparsification in dynamic streams; see~\cite{mcgregor2014graph} for a
survey of graph streaming. Kapralov et al.~\cite{KLM+17} extended this
approach to spectral sparsifiers, and Kapralov et al.~\cite{KMM+20}
reduced the recovery time to nearly linear. These algorithms handle
weighted graphs by grouping edges into geometric weight classes, which
requires each deletion to reveal the weight used at insertion. Chen,
Khanna, and Li~\cite{CKL22} gave the first linear sketch for weighted cut
sparsification in general turnstile streams, recovering strength-sampled
edges with $\ell_1$ samplers; see~\cite{JW21,LNSW25} for recent work on
$\ell_p$ sampling. Like the improved algorithm of~\cite{AGM12}, we read
sampled edges through cut queries on vertex sketches, with query cuts
from a Gomory--Hu tree of a coarse sparsifier. Unlike~\cite{AGM12}, we
must estimate the sampled weights, and we need only constant-accuracy
estimates.

\paragraph{Gomory--Hu trees and pseudorandomness.}
Gomory and Hu~\cite{GH61} showed that the pairwise minimum cuts of an
undirected graph are represented by a weighted tree, which can be
computed with $n-1$ maximum-flow computations. Li and
Panigrahi~\cite{LP21} computed approximate Gomory--Hu trees faster than
$n-1$ maximum flows, and Li, Nanongkai, Panigrahi, and
Saranurak~\cite{LNPS23} computed $(1+\eps)$-approximate Gomory--Hu trees
in near-linear time; our query cuts use the latter. Nisan~\cite{Nisan92}
gave pseudorandom generators for space-bounded computation, and Nisan
and Zuckerman~\cite{NZ96} gave a generator whose seed length is linear
in the space, for algorithms that use polynomially many random bits.
Kapralov et al.~\cite{KMM+20} combined the latter with a locally
computable extractor of De and Vidick~\cite{DV10}, which lets every
random bit of a dynamic-stream sketch be generated in polylogarithmic
time. We use their generator for the same purpose.

\paragraph{Lower bounds.}
Andoni et al.~\cite[Theorem~2.3]{ACK+15} showed that, for
$\eps>1/\sqrt n$, any data structure that approximately preserves all
cuts requires $\Omega(n/\eps^2)$ bits, and Carlson et al.~\cite{CKST19}
strengthened this to $\Omega(n\log n/\eps^2)$ bits for
$\eps=\omega(n^{-1/4})$. These bounds apply in particular to linear
sketches and turnstile streaming algorithms. Andoni et
al.~\cite[Theorem~2.4]{ACK+15} also showed that, for $\eps>1/\sqrt n$,
cut sparsifiers may need $\Omega(n/\eps^2)$ edges. As explained in
Section~\ref{sec:result}, these bounds make the space and the recovery
time of Theorem~\ref{thm:main} optimal up to polylogarithmic factors.

\section{Preliminaries}\label{sec:prelims}

\paragraph{Stream model and notation.}
We consider a fixed, oblivious turnstile stream on $n\ge2$ vertices.
There are $N=\binom n2$ coordinates, one for each unordered vertex pair,
in a fixed canonical order, which Section~\ref{sec:discovery} makes
explicit. The stream starts from zero and makes $M$
integer updates $(e,\Delta)$, each adding $\Delta$ to coordinate $e$,
with $|\Delta|\le U$ for some $U\ge1$. Intermediate weights may be
negative; final weights belong to $\{0,1,\ldots,U\}$. Let $G=(V,E,w)$ be
the final graph, with $E=\{e:w_e>0\}$. For disjoint vertex sets $S,T$,
write
\[
 E(S,T)=\{\{u,v\}\in E:u\in S,\ v\in T\},\qquad
 w(S,T)=\sum_{e\in E(S,T)}w_e.
\]
Thus $E(S,V\setminus S)$ is the set of positive-weight cut edges and
$w(S,V\setminus S)$ is the cut weight. For an edge set $F$, we also write
$w(F)=\sum_{e\in F}w_e$. For another graph $H$ or weight vector $x$
on the pairs, $w_H(S,T)$ and $x(S,T)$ sum the weights of $H$ or $x$ over
all pairs between $S$ and $T$, including pairs outside $E$; for
instance, $w_H(S,V\setminus S)$ is the cut weight in $H$. A \emph{positive reweighting} of edges of $G$ is a graph $H$
on $V$ with $E(H)\subseteq E$ and positive weights.

We assume $M,U,\delta^{-1}\le n^{O(1)}$, with $0<\eps<1$ and
$0<\delta<1/4$, where $\delta$ is the target failure probability. Thus
$\widetilde O(\cdot)$ hides factors polylogarithmic in $n$. We write
$[m]=\{1,\ldots,m\}$. A Rademacher sign is uniform on $\{-1,1\}$.

\paragraph{Model of computation.}
Running times refer to the word RAM with $\Theta(\log n)$-bit words;
under our assumptions, every input number fits in $O(1)$ words. The
algorithm knows $n$, $\eps$, $\delta$, $M$, and $U$ in advance. The
update time is the worst-case time to process one stream update, and the
recovery time is the worst-case time from the end of the stream until
the output is produced.

\begin{definition}[Cut sparsifier]\label{def:sparsifier}
For $0<\eps<1$, a weighted graph $H$ on $V$ is a $(1\pm\eps)$-cut
sparsifier of $G$ if
$(1-\eps)w(S,V\setminus S)\le w_H(S,V\setminus S)\le(1+\eps)w(S,V\setminus S)$
for every $S\subseteq V$.
\end{definition}

\subsection{Connectivity and cut counting}\label{sec:connectivity}

For a pair $e=uv$, its connectivity is the weight of the smallest cut
separating its endpoints:
\begin{equation}\label{eq:connectivity}
 \lambda_e=\min_{S:\,u\in S,\ v\notin S}w(S,V\setminus S).
\end{equation}
This is defined for every pair, including pairs with $w_e=0$. Every cut
separating the endpoints of $e$ has weight at least $\lambda_e$, and a
positive edge has $\lambda_e\ge w_e\ge1$. For another weighted graph $H$
or another weight vector $x$ on the pairs, $\lambda_H(e)$ and
$\lambda_e(x)$ denote the same quantity computed in $H$ or in $x$; we
also write $\lambda_H(u,v)=\lambda_H(e)$. Thus $\lambda_e=\lambda_e(w)$.
We use two facts from~\cite{FHHP11}.

\begin{lemma}[{\cite[Corollary~1.7]{FHHP11}}]\label{lem:sum-connectivity}
$\sum_{e\in E}w_e/\lambda_e\le n-1$.
\end{lemma}

\begin{lemma}[Projection counting]\label{lem:counting}
Let $G$ be connected, let $\lambda_0>0$, and let $F^\star\subseteq E$ be
a set of edges with $\lambda_e\ge\lambda_0$ for every $e\in F^\star$.
For every $\alpha\ge1$, the sets $E(S,V\setminus S)\cap F^\star$, over
all $\varnothing\ne S\subsetneq V$ with
$w(S,V\setminus S)\le\alpha\lambda_0$, take at most $n^{2\alpha}$
distinct values.
\end{lemma}
\begin{proof}
We use two results of~\cite{FHHP11}, which are stated for integer
weights, an edge of weight $w_e$ being treated as $w_e$ parallel unit
edges. This changes no cut value or connectivity, and parallel copies of
a pair lie in the same cuts, so projections in the multigraph determine
projections in $G$. Let $\lambda\ge1$ be the minimum cut of $G$. By
Karger's cut counting~\cite[Theorem~2.1]{FHHP11}, for every $\alpha\ge1$
at most $n^{2\alpha}$ cuts have value at most $\alpha\lambda$. Call an
edge $k$-heavy if its connectivity is at least $k$, and call the set of
$k$-heavy edges of a cut its $k$-projection. By the $k$-projection
theorem~\cite[Theorem~2.3]{FHHP11}, for every $k\ge\lambda$ and every
$\alpha\ge1$, the cuts of value at most $\alpha k$ have at most
$n^{2\alpha}$ distinct $k$-projections.

If $\lambda_0\le\lambda$, every counted cut has value at most
$\alpha\lambda_0\le\alpha\lambda$, so cut counting bounds even the number
of these cuts by $n^{2\alpha}$. If $\lambda_0>\lambda$, put
$k=\lceil\lambda_0\rceil$, so that $k\ge\lambda$. Connectivities are
integers, so every edge of $F^\star$ is $k$-heavy, and
$E(S,V\setminus S)\cap F^\star$ is determined by the $k$-projection of
the cut. A counted cut has value at most $\alpha\lambda_0\le\alpha'k$,
where $\alpha'=\max\{1,\alpha\lambda_0/k\}$ lies in $[1,\alpha]$. The
$k$-projection theorem with $k$ and $\alpha'$ therefore gives at most
$n^{2\alpha'}\le n^{2\alpha}$ distinct values.
\end{proof}

\subsection{Concentration inequalities}\label{sec:concentration}

We use the following standard bounds; see, for example,~\cite{BLM13}.

\begin{lemma}[Chernoff bound]\label{lem:chernoff}
Let $X$ be a sum of independent $[0,1]$-valued random variables, and let
$\mu\ge\E X$. For every $a>0$,
$\Pp(X\ge(1+a)\mu)\le\exp\{-a^2\mu/(2+a)\}$.
\end{lemma}

\begin{lemma}[Freedman's inequality~{\cite[Theorem~1.1]{Tropp11}}]
\label{lem:freedman}
Let $(Y_i)_{i\ge0}$ be a real-valued martingale with respect to a
filtration $(\mathcal H_i)_{i\ge0}$, with $Y_0=0$. Assume that its
differences $d_i=Y_i-Y_{i-1}$ satisfy $d_i\le D$ almost surely for all
$i\ge1$, and let
\[
 V_i=\sum_{j=1}^i\E\bigl[d_j^2\mid\mathcal H_{j-1}\bigr]
\]
be its predictable quadratic variation. Then for all $z\ge0$ and $v>0$,
\[
 \Pp\bigl(\exists i\ge0:\ Y_i\ge z\ \text{and}\ V_i\le v\bigr)
 \le\exp\!\left\{-\frac{z^2/2}{v+Dz/3}\right\}.
\]
\end{lemma}

We use the lemma at a fixed time $m$ and for both tails. If
$|d_i|\le D$ for all $i\in[m]$, then applying the lemma to $(Y_i)$ and
to $(-Y_i)$ shows that $X=Y_m$ and $V=V_m$ satisfy
\begin{equation}\label{eq:freedman}
 \Pp\bigl(|X|\ge z\ \text{and}\ V\le v\bigr)
 \le2\exp\!\left\{-\frac{z^2}{2(v+Dz/3)}\right\}.
\end{equation}
For every $\varrho>0$, the choice $z=\sqrt{2v\varrho}+\frac23D\varrho$
makes the right-hand side at most $2e^{-\varrho}$. Because the event
in~\eqref{eq:freedman} includes $V\le v$, a variance bound that holds
only on a good event can be used without conditioning the increments on
that event. For independent centered summands with their natural
filtration, $V$ is deterministic and~\eqref{eq:freedman} is Bernstein's
inequality. For $d_i=a_i\omega_i$, with $\mathcal H_{i-1}$-measurable
$a_i$ and an $\mathcal H_i$-measurable Rademacher sign $\omega_i$
independent of $\mathcal H_{i-1}$, we simply have $V=\sum_{i=1}^ma_i^2$.

\subsection{Gomory--Hu trees and a constant-accuracy sparsifier}
\label{sec:gh}

Let $C$ be the vertex set of a connected component of a nonnegatively
weighted graph $H$. A Gomory--Hu tree~\cite{GH61} of $H$ on $C$ is a
weighted tree on $C$ with the following property. Fix a root. For a tree
edge $q$, let $\tau_q$ be its weight, and let the fundamental cut $S_q$
be the vertex set of the component of the tree minus $q$ that contains
the root. Then $w_H(S_q,C\setminus S_q)=\tau_q$ for every tree edge $q$,
and for every pair $u,v\in C$, the connectivity $\lambda_H(u,v)$ equals
the minimum of $\tau_q$ over the edges $q$ on the $u$--$v$ tree path. In
particular, if $q$ is a minimum-weight edge on that path, then $S_q$ is a
minimum cut separating $u$ and $v$.

To obtain query cuts, we give a constant-accuracy sparsification
sketch with near-linear recovery time. The following theorem states its
guarantee, and Section~\ref{sec:coarse} proves it.

\begin{theorem}[Constant-accuracy sparsification]\label{thm:coarse}
In the stream model above, for $0<\delta_0<1/4$ with
$\delta_0^{-1}\le n^{O(1)}$, there is a one-pass linear sketch that uses
$\widetilde O(n)$ bits of space, including seeds and recovery workspace,
$\widetilde O(1)$ worst-case time per update, and $\widetilde O(n)$
worst-case recovery time. It always returns a graph with
$\widetilde O(n)$ edges, and with probability at least $1-\delta_0$ this
graph is a positive reweighting $H_0$ of edges of $G$ with
\begin{equation}\label{eq:coarse-accuracy}
 \frac{24}{25}w(S,V\setminus S)\le w_{H_0}(S,V\setminus S)
                    \le\frac{26}{25}w(S,V\setminus S)
                    \qquad(S\subseteq V).
\end{equation}
\end{theorem}

At constant accuracy, a polynomial-time sketch with this guarantee is
also available from Chen, Khanna, and
Li~\cite[Theorem~I.1 and Corollary~I.2]{CKL22}. Our proof
uses only the recovery procedure of this paper and the tools of
Section~\ref{sec:nearlinear}. The rest of the paper uses only the stated
guarantee.

\paragraph{Approximate Gomory--Hu trees.}
Section~\ref{sec:query-forest} replaces exact Gomory--Hu trees by
approximate ones. Following Li and Panigrahi~\cite{LP21}, a
$(1+\eps_0)$-approximate Gomory--Hu tree of a connected graph $H$ is a
weighted tree $T$ on the vertex set of $H$ with the following property.
For all vertices $u,v$, let $q$ be a minimum-weight edge on the $u$--$v$
path of $T$. Then the component of $T-q$ that contains $u$ is a cut of
$H$ of weight at most $(1+\eps_0)\lambda_H(u,v)$, and this weight equals
the weight of $q$. Such a cut separates $u$ and $v$, since $q$ lies on
their path.

\begin{theorem}[{\cite[Theorem~1.2]{LNPS23}}]\label{thm:approx-gh}
For every $\eps_0>0$, there is a randomized algorithm that, given a
connected undirected graph $H$ with $m_H$ edges and positive weights
whose ratio is polynomially bounded, computes a
$(1+\eps_0)$-approximate Gomory--Hu tree of $H$ in
$\widetilde O(m_H\operatorname{poly}(1/\eps_0))$ time with high
probability.
\end{theorem}

In the full version of~\cite{LNPS23}, this result is Theorem~1.4. As
usual, with high probability means with probability at least
$1-n_H^{-c}$, where $n_H$ is the number of vertices of $H$ and $c>0$ is
any constant fixed in advance. We use only the tree, not its weights.
Section~\ref{sec:query-forest} computes the weight of every fundamental
cut that it needs directly in the graph.

\subsection{Randomness}\label{sec:randomness}

Our sketches use two kinds of randomness. Hash functions and most signs
need only four-wise independence. We use the standard polynomial
construction over a binary extension field; projecting field elements
gives uniform power-of-two buckets and unbiased signs, each seed has
$O(\log n)$ bits, and each value takes polylogarithmic time to compute
(Appendix~\ref{app:precision}).

Two families, the survival bits and the evaluation signs of
Section~\ref{sec:tables}, are analyzed as fully independent, and storing
them explicitly would take $\Omega(n^2)$ bits. Section~\ref{sec:tapes}
generates both from short seeds with the following pseudorandom
generator for space-bounded computation. Kapralov et
al.~\cite{KMM+20} built it from the generator of Nisan and
Zuckerman~\cite{NZ96} and a locally computable extractor of De and
Vidick~\cite{DV10}. As usual for such
generators~\cite{Nisan92,NZ96}, an algorithm runs in space $S$ if it has
$S$ bits of working memory and read-only access to its input, and it
reads its random bits once, in a fixed order.

\begin{proposition}[{\cite[Theorem~4.1]{KMM+20}}]\label{prop:prg}
For any constants $q,c>0$ and every $S\ge2$, there is an explicit
generator that stretches a seed of $O(S\operatorname{polylog}S)$ random
bits into $S^q$ bits with the following property. If a randomized
algorithm runs in space $S$ and reads at most $S^q$ random bits, then
replacing these bits by the output
of the generator on a uniformly random seed increases its failure
probability by at most $S^{-c}$. Each output bit can be computed from
the seed in $O(\log^{O(q)}S)$ time.
\end{proposition}

Throughout Sections~\ref{sec:algorithm}--\ref{sec:allcuts}, survival
bits and evaluation signs are fully independent.

\section{The algorithm}\label{sec:algorithm}

During the stream, we maintain the sketch of Theorem~\ref{thm:coarse},
which we call the coarse sketch, and, independently, three tables at each
vertex. After the stream, we compute the coarse sparsifier $H_0$, build
its Gomory--Hu trees, and use their query cuts to select and reweight
edges. This section presents the polynomial-time version of the
algorithm, in which the decoder tests all vertex pairs and the query
cuts come from exact Gomory--Hu trees. Section~\ref{sec:nearlinear}
replaces both steps by near-linear-time implementations and generates
the random bits from short seeds.

\subsection{Parameters and query cuts}\label{sec:params}

We use $J+1$ sampling layers, $R$ independent rows per sketch, and $B$
buckets per row. Their roles are distinct: $J$ covers the weight range,
$R$ makes local failures unlikely, and $B$ controls collision noise.
Set
\begin{equation}\label{eq:params}
 J=\lceil\log_2(8n^4U)\rceil,\qquad
 \ell=\lceil\ln(64n^2J/\delta)\rceil,\qquad
 \beta=\lceil C_\beta J\ell/\eps^2\rceil.
\end{equation}
Let $R$ be the smallest odd integer at least $C_R\ell$, and let $B$ be
the smallest power of two at least $C_B\beta$. We also use $R$ for the
number of evaluation repetitions and diagnostic groups. Choose
sufficiently large absolute constants $C_B$ and $C_R$, and then
$C_\beta$. In particular, $J,R=O(\log n)$, $B=O(\beta)$, and
$\beta=O(\eps^{-2}\log^2n)$. Table~\ref{tab:params} summarizes these
choices.

\begin{table}[ht]
\centering
\begin{tabular}{@{}lll@{}}
\toprule
Symbol & Value & Role\\
\midrule
$J$ & $\lceil\log_2(8n^4U)\rceil$ & layers $0,\ldots,J$ cover the weight range\\
$\ell$ & $\lceil\ln(64n^2J/\delta)\rceil$ & logarithmic confidence parameter\\
$\beta$ & $\lceil C_\beta J\ell/\eps^2\rceil$ & sampling budget\\
$R$ & smallest odd integer $\ge C_R\ell$ & rows, sign repetitions, diagnostic groups\\
$B$ & smallest power of two $\ge C_B\beta$ & buckets per row; controls collision noise\\
$d$ & $\lceil\log_2N\rceil$ & bits of a pair identifier\\
$B_S$ & smallest power of two $\ge C_B\beta(d+1)$ & selection buckets per row\\
\bottomrule
\end{tabular}
\caption{Parameters of the main algorithm. The last two are used only
from Section~\ref{sec:discovery} on.}
\label{tab:params}
\end{table}

\paragraph{The coarse graph.}
Run the sketch of Theorem~\ref{thm:coarse} independently of the three
tables below, with $\delta_0=\delta/20$. Let $H_0$ be its output, and
write $s_0=|E(H_0)|$. By Theorem~\ref{thm:coarse}, $s_0=\widetilde O(n)$
on every run.

\paragraph{The query cuts.}
Build an exact Gomory--Hu tree in each nontrivial component of $H_0$,
with a fixed root. A tree edge $q$ has weight $\tau_q$ and fundamental
cut $S_q$ as in Section~\ref{sec:gh}. For a pair $e=uv$ whose endpoints
lie in a common nontrivial component, choose a minimum-weight edge $q(e)$
on the $u$--$v$ path of its tree and set
\begin{equation}\label{eq:query-cut}
 \kappa_e=\tau_{q(e)},\qquad S_e=S_{q(e)},\qquad
 t_e(k)=2^{-k}\kappa_e/\beta\quad(k\ge0).
\end{equation}
We call $S_e$ the query cut of $e$, $\kappa_e$ its query weight, and
$t_e(k)$ its scale at layer $k$. A pair is \emph{eligible} if its endpoints belong to the
same connected component of $H_0$; all other pairs are skipped. All ties
are broken deterministically.

If $H_0$ satisfies~\eqref{eq:coarse-accuracy}, then a cut has weight
zero in $H_0$ exactly when it has weight zero in $G$, so $H_0$ and $G$
have the same connected components. For every eligible pair $e$, the
Gomory--Hu property gives
$\kappa_e=\lambda_{H_0}(e)=w_{H_0}(S_e,V\setminus S_e)$, and together
with~\eqref{eq:coarse-accuracy} this yields
\begin{equation}\label{eq:calibration}
 .9\lambda_e\le\kappa_e\le1.1\lambda_e,\qquad
 |e\cap S_e|=1,\qquad
 w(S_e,V\setminus S_e)\le\kappa_e/.9.
\end{equation}
The middle condition says that $S_e$ separates the endpoints of $e$,
including when $w_e=0$. The weight of $S_e$ is within a constant factor
of $\lambda_e$. The trees provide at most $n-1$ distinct query cuts.
Section~\ref{sec:query-forest} obtains query cuts with the same
guarantees in near-linear time from an approximate Gomory--Hu tree.

\subsection{The three tables maintained at each vertex}\label{sec:tables}

Because the query cuts are determined after the stream, we maintain
sketches of signed incidence vectors at the vertices. Orient each pair
from its smaller endpoint to its larger endpoint. The signed incidence
vector $\chi_v\in\mathbb R^N$ of a vertex $v$ has $\chi_v(e)=w_e$ if $v$
is the smaller endpoint of $e$, $\chi_v(e)=-w_e$ if $v$ is the larger
endpoint, and $\chi_v(e)=0$ otherwise. Then
\[
 \chi_S:=\sum_{v\in S}\chi_v,\qquad
 \chi_S(e)=\sgn_S(e)\,w_e,\qquad
 \sgn_S(e)=\ind[\min e\in S]-\ind[\max e\in S].
\]
Summing vertex sketches cancels internal edges and leaves exactly a
signed sketch of the cut. This cancellation requires all vertices to use
the same hashes and signs for a given table, layer, and row. Write
$\sgn_e=\sgn_{S_e}(e)$; by~\eqref{eq:calibration}, $\sgn_e\in\{-1,1\}$.

Choose fair bits $\xi_{e,a}\in\{0,1\}$ independently over all
coordinates $e$ and layers $a\in[J]$, and define the survival indicator
\[
 I_{e,k}=\prod_{a=1}^k\xi_{e,a},\qquad I_{e,0}=1.
\]
We say that $e$ survives to layer $k$ if $I_{e,k}=1$. The samples are
nested, and $\Pp(I_{e,k}=1)=2^{-k}$. Every update to a coordinate uses
the same bits and sketch coefficients.

Each vertex $v$ maintains the following tables at each layer
$k\in\{0,\ldots,J\}$. Every table has $R$ rows and $B$ buckets per row.
\begin{center}
\begin{tabular}{@{}llll@{}}
\toprule
Table & Hash in row $r$ & Counters per bucket & Purpose\\
\midrule
Selection $\Sel$ & $h_S^{k,r}$ & $1$ & choose the edge and layer\\
Diagnostic $\Diag$ & $h_T^{k,r}$ & $32R$ pairs & measure collision noise\\
Estimation $\Est$ & $h_T^{k,r}$ & $R$ & estimate the weight\\
\bottomrule
\end{tabular}
\end{center}
Different rows independently hash the same sampled edges into buckets.
Counters within a row keep the buckets fixed and vary only the signs.
We call the pair that the decoder currently queries the target; each
target will choose its own row for estimation.

The selection hashes $h_S^{k,r}:[N]\to[B]$ and estimation hashes
$h_T^{k,r}:[N]\to[B]$ are independent families. Let $s_e^{k,r}$ be the
selection signs, $\rho_e^{k,r,p,i}$ the diagnostic signs ($p\in[32R]$,
$i\in\{1,2\}$), and $\sigma_e^{k,r,l}$ the evaluation signs ($l\in[R]$).
The counters are
\begin{align}
 \Sel^{k,r}(v,b)
 &=\sum_{e:\,h_S^{k,r}(e)=b}\chi_v(e)\,I_{e,k}\,s_e^{k,r},\label{eq:S}\\
 \Diag^{k,r,p,i}(v,b)
 &=\sum_{e:\,h_T^{k,r}(e)=b}\chi_v(e)\,I_{e,k}\,\rho_e^{k,r,p,i},\label{eq:D}\\
 \Est^{k,r,l}(v,b)
 &=\sum_{e:\,h_T^{k,r}(e)=b}\chi_v(e)\,I_{e,k}\,\sigma_e^{k,r,l}.
 \label{eq:T}
\end{align}
The diagnostic and estimation counters use the same hash, so they
contain the same colliding edges. All three tables use the same survival
indicators. The hashes, selection signs, and diagnostic signs are
four-wise independent across coordinates, with independent seeds for
distinct hash functions and sign families. These seeds are independent
of the survival bits and of the coarse sketch. The evaluation signs are
fully independent of one another and of all other randomness.

An update $(e,\Delta)$ with $e=uv$ and $u<v$ changes only the tables at
$u$ and $v$. At each layer $k$ with $I_{e,k}=1$, add $\Delta$ times the
appropriate sign to the appropriate bucket at $u$, and subtract it at
$v$, for every row and sign repetition. Hashes, signs, and survival bits
are reused for every update to $e$. Thus these are linear sketches
throughout the stream, including when intermediate edge weights are
negative.

\subsection{Recovery: select, diagnose, and estimate}\label{sec:recovery}

After computing $H_0$ and its trees, sum the vertex counters on each
fundamental cut. For any table, replacing the vertex argument $v$ by a
set $S$ means summing over $v\in S$; for example,
$\Sel^{k,r}(S,b)=\sum_{v\in S}\Sel^{k,r}(v,b)$. These are sketches of
the sampled cut edges, with internal edges cancelled. Tree aggregation
takes one postorder traversal for all counter arrays; see
Section~\ref{sec:discovery}.

\paragraph{Select the edge and layer.}
Enumerate eligible vertex pairs one at a time;
Section~\ref{sec:discovery} replaces this enumeration by candidate
discovery. For a pair $e$, the query
\[
 y_e(k)=\med_{r\in[R]}\
   \sgn_e\,s_e^{k,r}\,\Sel^{k,r}\bigl(S_e,h_S^{k,r}(e)\bigr)
\]
estimates $w_eI_{e,k}$. The decoder compares it with the scale $t_e(k)$
and takes the first layer that passes:
\begin{equation}\label{eq:selection}
 \theta_e(k)=\max\{1/2,\,t_e(k)/4\},\qquad
 \widehat j_e=\min\{0\le k\le J:\ y_e(k)\ge\theta_e(k)\}.
\end{equation}
An empty minimum is infinity. The floor $1/2$ uses the fact that
positive weights are at least one; it prevents tiny noisy queries from
accepting zero coordinates. Let $E_\star$ be the pairs with finite
$\widehat j_e$. Store only this list, and abort if it exceeds
$\lceil8n\beta\rceil$ pairs.

\paragraph{Choose a row with small collision noise.}
For each $e\in E_\star$, put $k=\widehat j_e$. The estimation table at
layer $k$ has $R$ rows with independent hashes, so $e$ meets different
colliding edges in different rows. The decoder chooses one of these $R$
rows, separately for each selected edge, and uses only this row to
estimate $w_e$. It makes this choice without knowing the target weight
or using the evaluation signs. In row $r$, query the target bucket
$b=h_T^{k,r}(e)$ with each pair of diagnostic signs and subtract:
\[
 Y_e^{r,p}=\frac{\sgn_e}{2}
 \Bigl(\rho_e^{k,r,p,1}\,\Diag^{k,r,p,1}(S_e,b)
      -\rho_e^{k,r,p,2}\,\Diag^{k,r,p,2}(S_e,b)\Bigr).
\]
We call $Y_e^{r,p}$ a \emph{difference query}. Each signed query has the
same target contribution $w_eI_{e,k}$, which cancels in the difference.
Thus the squared difference query measures only collision noise.
Partition the $32R$ difference queries into $R$ groups
$P_1,\ldots,P_R$ of size $32$, and set
\begin{equation}\label{eq:diagnostic}
 \psi_r(e)=\med_{g\in[R]}\frac1{32}\sum_{p\in P_g}\bigl(Y_e^{r,p}\bigr)^2,
 \qquad
 r_e=\arg\min_{r\in[R]}\psi_r(e).
\end{equation}
Break ties deterministically. Section~\ref{sec:diagnostic} shows that
$\psi_r(e)$ estimates the collision energy of row $r$ within constant
factors, so minimizing it finds a row with small collision noise.

\paragraph{Estimate the weight with independent signs.}
In the chosen row $r_e$, read the $R$ estimation counters of the target
bucket, average their signed queries, and undo the subsampling:
\begin{equation}\label{eq:output}
 \widehat w_e=\frac1R\sum_{l=1}^R
  \sgn_e\,\sigma_e^{k,r_e,l}\,
  \Est^{k,r_e,l}\bigl(S_e,h_T^{k,r_e}(e)\bigr),\qquad
 Z_e=2^k\widehat w_e.
\end{equation}
Here $\widehat w_e$ estimates the original weight; $Z_e$ also
compensates for sampling at rate $2^{-k}$. Return $H$ with edges
$E_\star$ and weights $Z_e$, aborting if any weight is nonpositive.

All edge, layer, and row choices, including the check on the number of
selected pairs, precede
the use of evaluation signs in recovery. Conditional on these choices,
the estimate of a surviving target is unbiased, and its error has the
quadratic-sign form analyzed in Section~\ref{sec:conditional}. The
decoder chooses layers solely through sketch queries and never tests
survival bits directly; this also permits the seed implementation in
Section~\ref{sec:tapes}.

\section{Local sketch guarantees}\label{sec:local}

Sections~\ref{sec:local}--\ref{sec:allcuts} analyze the recovery
procedure of Section~\ref{sec:recovery} for fixed query cuts. Recall from
Section~\ref{sec:params} that every eligible pair $e$ has a query cut
$S_e$, which separates its endpoints, and a query weight $\kappa_e$,
which estimates $\lambda_e$ and sets the scales $t_e(k)$. The analysis
uses only the following two properties, and not how the query cuts are
computed.

\begin{definition}[Accurate query cuts]\label{def:accurate}
The query cuts are \emph{accurate} if the eligible pairs are exactly the
pairs $e$ with $\lambda_e>0$, and every eligible pair $e$
satisfies~\eqref{eq:calibration}.
\end{definition}

Exact Gomory--Hu trees of $H_0$ give accurate query cuts whenever $H_0$
satisfies~\eqref{eq:coarse-accuracy} (Section~\ref{sec:params}), and so
do the query forests of Section~\ref{sec:query-forest}. In both cases,
the query cuts depend only on the coarse sketch and on randomness of
their own, which are independent of the three tables. Throughout
Sections~\ref{sec:local}--\ref{sec:allcuts}, we therefore fix accurate
query cuts, and all probabilities are over the randomness of the three
tables.

We first show that the selection queries are accurate
(Section~\ref{sec:selection}) and that accurate selection queries
restrict each positive edge to two possible layers
(Section~\ref{sec:twostep}). We then show that the diagnostic finds a
row with small collision noise (Section~\ref{sec:diagnostic}), which
gives small, centered weight-estimation errors
(Section~\ref{sec:estimate}). Each local failure probability is at most
$\delta/20$, with the constants in~\eqref{eq:params}, conditional on the
prerequisites stated in the corresponding lemma.

\subsection{Uniform accuracy of the selection queries}\label{sec:selection}

For a query cut of weight $O(\kappa_e)$, subsampling reduces the
expected total weight to $O(2^{-k}\kappa_e)$. Dividing by the scale
$t_e(k)=2^{-k}\kappa_e/\beta$ leaves $O(\beta)$. The following load bound
makes this observation uniform over all query cuts and layers.

\begin{lemma}[Load and point queries]\label{lem:selection}
For accurate query cuts, with failure probability at most $\delta/20$ over
the survival bits (the fair coins $\xi_{f,a}$ that determine which pairs
survive to which layers; see Section~\ref{sec:tables}),
\begin{equation}\label{eq:load}
 \sum_{f\in E(S_e,V\setminus S_e)}
 \min\{w_f/t_e(k),1\}\,I_{f,k}\le2.5\beta
\end{equation}
for every eligible pair $e$ and every layer $k$. Fix any survival
bits satisfying this bound. Then, with additional failure probability
at most $\delta/20$ over the selection seeds,
\begin{equation}\label{eq:pointgood}
 |y_e(k)-w_eI_{e,k}|\le t_e(k)/24
\end{equation}
for every eligible pair and layer, including zero coordinates.
\end{lemma}
\begin{proof}
Each load is a sum of independent $[0,1]$-valued variables. Since
$\E I_{f,k}=2^{-k}$ and $t_e(k)=2^{-k}\kappa_e/\beta$, its mean is at most
$\sum_fw_f\beta/\kappa_e=\beta\,w(S_e,V\setminus S_e)/\kappa_e\le(10/9)\beta$
by~\eqref{eq:calibration}.
Lemma~\ref{lem:chernoff} and a union bound over the $N(J+1)$ loads
give~\eqref{eq:load}.

Fix an eligible pair $e$ and a layer $k$, and write $t=t_e(k)$. Edges
that do not cross $S_e$ cancel. Call a surviving edge crossing $S_e$
heavy if its weight exceeds $t$, and light otherwise. The load bound
gives at most $2.5\beta$ heavy survivors, and light survivors of total
squared weight at most $2.5\beta t^2$. In one row, the signed reading of
the bucket of $e$ is $w_eI_{e,k}$ plus the contributions of the other
survivors in that bucket. The target collides with another heavy
survivor with probability at most $2.5\beta/B$. The noise from other
light survivors has second moment at most $2.5\beta t^2/B$, so its
probability of exceeding $t/24$ is at most $1440\beta/B$. For large
enough $C_B$, a row is therefore accurate, that is, its reading is within
$t/24$ of $w_eI_{e,k}$, with probability at least $.95$. Taking the median of $R$
independent rows and a union bound over the $N(J+1)$ queries prove the
claim.
\end{proof}

We refer to~\eqref{eq:pointgood}, for all eligible pairs and layers, as
\emph{selection success}.

\subsection{Reduction to two sampling layers}\label{sec:twostep}

Recall from Section~\ref{sec:recovery} that the decoder selects a pair
$e$ at the first layer $\widehat j_e$ at which its selection query
$y_e(k)$ reaches the threshold $\theta_e(k)=\max\{1/2,\,t_e(k)/4\}$, and
that $E_\star$ is the set of selected pairs, namely those with finite
$\widehat j_e$. This section shows that each positive edge can be
selected at only one of two consecutive layers, and that the choice
between them is made before the survival coin of $e$ that separates
them is read.

\paragraph{The baseline layer.}
Associate each positive edge with the baseline layer
\begin{equation}\label{eq:baseline-layer}
 k_e=\min\{k\ge0:\ w_e\ge5t_e(k)/24\},
\end{equation}
where $t_e(k)=2^{-k}\kappa_e/\beta$ is defined for all integers $k\ge0$.
The decoder does not compute $k_e$: it depends on the unknown final
weight. Once $G$ and the query cuts are fixed, however, $k_e$ is
deterministic. Since $w_e\ge1$ and $t_e(0)\le1.1(n-1)U$, we have
$k_e+1\le J$.

\paragraph{The hypothetical query.}
As in the proof of Lemma~\ref{lem:selection}, the signed reading of the
bucket of $e$ in one row is $w_eI_{e,k}$ plus the contributions of the
other surviving pairs in that bucket. A median commutes with adding a
constant, so
\[
 y_e(k)=w_eI_{e,k}+Z_e(k),
\]
where $Z_e(k)$, the median over the $R$ rows of these collision
contributions, depends on the selection seeds and on the survival bits
of the other pairs, but not on the survival bits of $e$. We call
\[
 y_e^+(k)=w_e+Z_e(k)=y_e(k)+w_e(1-I_{e,k})
\]
the \emph{hypothetical query} of $e$ at layer $k$: it is the selection
query that the decoder would see if $e$ survived to layer $k$ and all
other pairs were unchanged. The decoder never computes it. We use it
only to define
\begin{equation}\label{eq:twostep}
 \eta_e=\ind\bigl[y_e^+(k_e)<\theta_e(k_e)\bigr],\qquad j_e=k_e+\eta_e.
\end{equation}
Thus $\eta_e$ records whether the hypothetical query fails at the
baseline layer. Because $y_e^+$ does not involve the survival bits of
$e$, neither does $\eta_e$. Defining $\eta_e$ by the actual query
$y_e(k_e)$ instead would make it equal to one whenever $e$ did not
survive to layer $k_e$, so the layer would depend on the survival of
$e$ itself.

\begin{lemma}[Two possible layers]\label{lem:twostep}
Assume selection success~\eqref{eq:pointgood}. Then:
\begin{enumerate}
\item a nonedge is never selected;
\item a positive edge $e$ is selected if and only if it survives to
layer $j_e$, and in that case it is selected at layer $j_e$; that is,
\[
 e\in E_\star\ \Longleftrightarrow\ I_{e,j_e}=1,\qquad
 \widehat j_e=j_e\ \text{ if }e\in E_\star;
\]
\item if $\eta_e=1$, then $w_e<7t_e(k_e)/24$, that is,
$w_e2^{k_e}<7t_e(0)/24$.
\end{enumerate}
\end{lemma}
\begin{proof}
Fix a layer $k$ and write $t=t_e(k)$. By~\eqref{eq:pointgood},
$|Z_e(k)|=|y_e(k)-w_eI_{e,k}|\le t/24$, so
$|y_e^+(k)-w_e|\le t/24$. Moreover, the actual query satisfies
$y_e(k)=y_e^+(k)$ if $I_{e,k}=1$, and $y_e(k)=Z_e(k)\le t/24<\theta_e(k)$
if $I_{e,k}=0$. Hence the actual query passes at layer $k$ if and only
if $e$ survives to layer $k$ and the hypothetical query passes there.
For a nonedge, $y_e(k)=Z_e(k)$ at every layer, so it is never selected.

Next, the hypothetical query of a positive edge first passes at layer
$j_e$. For $k<k_e$, the definition of $k_e$ gives $w_e<5t/24$, so
$y_e^+(k)<5t/24+t/24=t/4\le\theta_e(k)$. At layer $k_e$ it passes if and
only if $\eta_e=0$, by definition. At layer $k_e+1$, the scale has
halved: writing $t'=t_e(k_e+1)$, the definition of $k_e$ gives
$w_e\ge5t_e(k_e)/24=10t'/24$. The hypothetical query is therefore at
least $w_e-t'/24\ge.9w_e>1/2$, and also at least $9t'/24>t'/4$, so it
passes.

Now combine the two facts. At layers before $j_e$ the hypothetical
query fails, so the actual query fails as well. If $I_{e,j_e}=1$, the
actual query passes at $j_e$, so $\widehat j_e=j_e$. If $I_{e,j_e}=0$,
then by nested survival $I_{e,k}=0$ for every $k\ge j_e$, so the actual
query never passes and $e\notin E_\star$.

Finally, suppose $\eta_e=1$, and write $t=t_e(k_e)$. If $t\le2$, then
$\theta_e(k_e)=1/2$ and $y_e^+(k_e)\ge1-2/24>1/2$, contradicting
$\eta_e=1$. Hence $t>2$, the threshold is $t/4$, and failure gives
$w_e-t/24<t/4$, that is, $w_e<7t/24=7\cdot2^{-k_e}t_e(0)/24$.
\end{proof}

The quantity $\eta_e$ is determined by the selection seeds and the
survival bits $\xi_{f,a}$ of all pairs $f$ with $a\le k_e$. The coin
$\xi_{e,k_e+1}$, which decides whether $e$ survives from layer $k_e$ to
layer $k_e+1$, is independent of all of them. This is the only
adaptivity property needed in the sampling analysis
(Section~\ref{sec:extra}).

\subsection{The diagnostic finds a good row}\label{sec:diagnostic}

For a pair $e$, a layer $k$, and a row $r$, the collision vector records
the other sampled edges that enter the target's estimation bucket:
\[
 v_e^{k,r}(f)=\sgn_e\,\sgn_{S_e}(f)\,w_f\,I_{f,k}\,
     \ind\bigl[h_T^{k,r}(f)=h_T^{k,r}(e)\bigr]\quad(f\ne e),
 \qquad v_e^{k,r}(e)=0.
\]
The sign $\sgn_{S_e}(f)$ already removes pairs that do not cross $S_e$.
For a selected edge, write $v_e=v_e^{j_e,r_e}$ for the collision vector
of the row chosen by~\eqref{eq:diagnostic}.

\begin{lemma}[Diagnostic]\label{lem:diagnostic}
Fix accurate query cuts, and fix any survival bits and selection seeds
such that~\eqref{eq:load} and~\eqref{eq:pointgood} hold. Then, with
failure probability at most $\delta/20$ over the estimation hashes and
diagnostic signs, which are independent of the fixed choices, every
selected edge satisfies $\norm{v_e}_2\le t_e(j_e)/18$.
\end{lemma}

The argument has two parts: independent hashes provide a good row, and
the difference queries recognize one without knowing the target weight or
inspecting the evaluation signs.

\begin{proof}
\emph{A good row exists.} Fix an eligible pair $e$ and a layer $k$, and
write $t=t_e(k)$. The load bound gives at most $2.5\beta$ heavy
survivors crossing $S_e$, and light survivors of total squared weight at
most $2.5\beta t^2$.
For one estimation hash, the probability of a collision with another
heavy survivor is at most $2.5\beta/B$. By Markov's inequality, the
probability that the light collision squared mass exceeds $(t/36)^2$ is
also $O(\beta/B)$. Thus, for sufficiently large $C_B$, a row $r$ has
$\norm{v_e^{k,r}}_2\le t/36$ with probability at least $9/10$. The rows
are independent, so the probability that none is good is at most
$10^{-R}$. A union bound over the $N(J+1)$ target--layer pairs gives a
good row simultaneously for all of them.

\emph{Difference queries recognize a good row.} Fix the hashes, fix $e$, $k$,
and $r$, and write $v=v_e^{k,r}$. Consider one difference query $Y_e^{r,p}$
computed at layer $k$, and suppress the indices $k,r,p$ of its signs.
Since $\sgn_e^2=1$, the signed query with copy $i\in\{1,2\}$ equals
$w_eI_{e,k}+Z_i$, where $Z_i=\sum_fv(f)\rho_e^i\rho_f^i$. Hence
$Y_e^{r,p}=\frac12(Z_1-Z_2)$. The two copies are independent. Within a
copy, four-wise independence gives $\E Z_i=\E Z_i^3=0$,
$\E Z_i^2=\norm v_2^2$, and $\E Z_i^4\le3\norm v_2^4$: the factor
$\rho_e^i$ cancels in even powers, and $v(e)=0$. Consequently, the difference query
$Y=Y_e^{r,p}$ satisfies
\[
 \E Y^2=\tfrac12\norm v_2^2,\qquad
 \E Y^4\le\tfrac34\norm v_2^4,\qquad
 \Var(Y^2)\le\tfrac12\norm v_2^4.
\]
Distinct difference queries use independent sign families, so an
average of $32$ squared difference queries has variance at most $\norm v_2^4/64$. Cantelli's
inequality bounds the probabilities of falling below $\norm v_2^2/4$ and
above $\norm v_2^2$ by $1/5$ and $1/17$, respectively. Taking the median
over the $R$ independent groups and a union bound over all
$N(J+1)R$ triples $(e,k,r)$ give
\[
 \tfrac14\norm{v_e^{k,r}}_2^2\le\psi_r(e)\le\norm{v_e^{k,r}}_2^2
\]
simultaneously, where $\psi_r(e)$ is computed at layer $k$. Zero vectors
have zero diagnostics exactly.

\emph{Conclusion.} By Lemma~\ref{lem:twostep}, a selected edge has
$\widehat j_e=j_e$, so its diagnostic is computed at layer $k=j_e$.
Minimizing $\psi_r(e)$ then chooses a row whose squared collision norm
is at most $4\psi_{r_e}(e)\le4\min_r\psi_r(e)\le4(t/36)^2$, that is, a
norm at most twice the best row norm, and hence at most $t/18$. Choosing
$C_R$ large enough makes the combined failure probability of both parts
at most $\delta/20$.
\end{proof}

\subsection{Estimating a selected weight}\label{sec:estimate}

\begin{lemma}[Weight estimation]\label{lem:estimate}
Fix accurate query cuts, and fix any survival bits, selection seeds,
estimation hashes, and diagnostic signs such that~\eqref{eq:load},
\eqref{eq:pointgood}, and the conclusion of Lemma~\ref{lem:diagnostic}
hold. Conditional on these choices, every selected edge satisfies
\begin{equation}\label{eq:edgeerror}
 \widehat w_e-w_e=\frac1R\sum_{l=1}^R\sum_{f\ne e}
 v_e(f)\,\sigma_e^{j_e,r_e,l}\,\sigma_f^{j_e,r_e,l},
 \qquad \E_\sigma\widehat w_e=w_e.
\end{equation}
With additional failure probability at most $\delta/20$ over the
evaluation signs, $\widehat w_e>w_e/2>0$ for every selected edge.
\end{lemma}
\begin{proof}
By Lemma~\ref{lem:twostep}, a selected edge is a positive edge with
$\widehat j_e=j_e$ and $I_{e,j_e}=1$. Its self-contribution
to~\eqref{eq:output} is therefore $w_e$; the other terms
give~\eqref{eq:edgeerror}. The evaluation signs remain independent, so
the error is centered. Each repetition is subgaussian with variance
proxy $\norm{v_e}_2^2$. Since $j_e\ge k_e$, the baseline
definition~\eqref{eq:baseline-layer} gives
$t_e(j_e)\le t_e(k_e)\le24w_e/5$, and hence
$\norm{v_e}_2\le t_e(j_e)/18\le4w_e/15$. Averaging $R$ repetitions gives
$\widehat w_e>w_e/2$ for all selected edges by a union bound.
\end{proof}

Let $\mathcal E_{\rm pre}$ denote the event that~\eqref{eq:load},
\eqref{eq:pointgood}, and the conclusion of Lemma~\ref{lem:diagnostic}
hold, for the fixed accurate query cuts. This event involves no evaluation
signs. After fixing any realization in $\mathcal E_{\rm pre}$, the
coefficients in~\eqref{eq:edgeerror} are deterministic. Errors for
different edges still share signs; they are centered but need not be
independent. The next two sections control their sum without requiring
an $\eps$-accurate estimate of any individual weight.

\section{A conditional bound for shared random signs}\label{sec:conditional}

By~\eqref{eq:edgeerror}, the error of a selected edge $e$ is the
average over $l\in[R]$ of its own evaluation sign
$\sigma_e^{j_e,r_e,l}$ times its collision noise
$\sum_fv_e(f)\,\sigma_f^{j_e,r_e,l}$. The error on a cut is therefore a
quadratic form in the evaluation signs, and the errors of different
edges are not independent: the sign that decodes one edge may appear in
the collision noise of another. This section proves a tail bound for
such quadratic forms that holds simultaneously for exponentially many
sets of edges. Section~\ref{sec:est} applies it within each connectivity
class, with the selected edges of the class as targets,
$\pi_e=(j_e,r_e)$, $u_e=2^{j_e}v_e$, and the projections of cuts onto
the class (Section~\ref{sec:classes}) as $\mathcal P$.

\begin{lemma}[Conditional estimation bound]\label{lem:estimation}
Let the evaluation signs $\sigma_f^{\pi,l}$, indexed by a pair
$f\in[N]$, a layer and row $\pi$, and a repetition $l\in[R]$, be
independent Rademacher signs, and let all other quantities below be
fixed. Let $\mathcal T\subseteq[N]$ be a set of targets.
Each target $e$ is decoded at a layer and row $\pi_e$, with decoding
signs $\sigma_e^{\pi_e,l}$ for $l\in[R]$, and has a collision vector
$u_e\in\mathbb R^N$. Assume that targets decoded at the same layer and
row do not appear in each other's collision vectors:
\begin{equation}\label{eq:sign-separation}
 u_e(f)=0\qquad\text{for all }e,f\in\mathcal T\text{ with }\pi_e=\pi_f.
\end{equation}
In particular, $u_e(e)=0$. Define the collision noise of $e$ in
repetition $l$ and the error on a set $F\subseteq[N]$ by
\[
 c_{e,l}=\sum_fu_e(f)\,\sigma_f^{\pi_e,l},
 \qquad
 Q_F=\frac1R\sum_{e\in F\cap\mathcal T}\sum_{l=1}^R
                      \sigma_e^{\pi_e,l}\,c_{e,l}.
\]
Let $\mathcal P$ be a family of subsets of $[N]$, and for each
$F\in\mathcal P$ let $\ell_F>0$ and
$v_F\ge\sum_{e\in F\cap\mathcal T}\norm{u_e}_2^2$. Then
\begin{equation}\label{eq:estimation-tail}
 \Pp\Bigl(\exists F\in\mathcal P:\
       |Q_F|>\sqrt{18v_F\ell_F/R}\Bigr)
 \le|\mathcal T|e^{-R}+2\sum_{F\in\mathcal P}e^{-\ell_F}.
\end{equation}
\end{lemma}

\begin{proof}
A decoding sign $\sigma_f^{\pi_f,l}$ of a target $f$ can occur only in
the collision noise $c_{e,l}$ of targets $e$ with $\pi_e=\pi_f$, and
there its coefficient is $u_e(f)=0$ by~\eqref{eq:sign-separation}.
Hence all collision noise depends only on the other signs, and we
condition on them. We first show that a single event, which does not
depend on $F$, bounds the energy $\sum_lc_{e,l}^2$ of every target.
Given this event, each $Q_F$ is a linear combination of independent
decoding signs whose coefficients have bounded energy.

\emph{Step 1: energy of the collision noise.}
Fix a target $e$ with $u_e\ne0$ and write $u=u_e$. Each $c_{e,l}$ is a
sum of independent terms $u(f)\sigma_f^{\pi_e,l}$, and a Rademacher sign
$\sigma$ satisfies $\E e^{x\sigma}=\cosh x\le e^{x^2/2}$ for every real
$x$. Hence $\E e^{sc_{e,l}}\le e^{s^2\norm u_2^2/2}$ for every real $s$.
With an independent standard Gaussian $g$ and the identity
$\E_ge^{zg}=e^{z^2/2}$, this gives
\[
 \E\exp\!\left\{\frac{c_{e,l}^2}{4\norm u_2^2}\right\}
 =\E_g\E_\sigma\exp\!\left\{\frac{g\,c_{e,l}}{\sqrt2\norm u_2}\right\}
 \le\E_ge^{g^2/4}=\sqrt2.
\]
The collision noises $c_{e,1},\dots,c_{e,R}$ involve disjoint sets of
signs and are therefore independent, so
$\E\exp\{\sum_lc_{e,l}^2/(4\norm u_2^2)\}\le2^{R/2}$. By Markov's
inequality,
\[
 \Pp\Bigl(\sum_lc_{e,l}^2>9R\norm u_2^2\Bigr)\le2^{R/2}e^{-9R/4}\le e^{-R}.
\]
If $u_e=0$, every $c_{e,l}$ vanishes. A union bound over targets
therefore shows that
\begin{equation}\label{eq:coefficient-energy}
 \sum_{l=1}^Rc_{e,l}^2\le9R\norm{u_e}_2^2\qquad\text{for all }e\in\mathcal T
\end{equation}
fails with probability at most $|\mathcal T|e^{-R}$. This needs no
independence between the collision noise of different targets.

\emph{Step 2: the decoding signs.}
Fix the other signs at any values for
which~\eqref{eq:coefficient-energy} holds. The collision noise is then
fixed, and $Q_F$ is a linear combination of the decoding signs
in which each sign appears at most once. Writing $\E_{\rm dec}$ for
expectation over the decoding signs, we have for every real $s$
\[
 \E_{\rm dec}e^{sQ_F}
 =\prod_{e\in F\cap\mathcal T}\prod_{l=1}^R\cosh(s\,c_{e,l}/R)
 \le\exp\!\left\{\frac{s^2}{2R^2}
                  \sum_{e\in F\cap\mathcal T}\sum_lc_{e,l}^2\right\}
 \le\exp\!\left\{\frac{9s^2v_F}{2R}\right\}.
\]
By Chernoff's method for both tails, $|Q_F|>\sqrt{18v_F\ell_F/R}$ with
conditional probability at most $2e^{-\ell_F}$. A union bound over
$F\in\mathcal P$, together with the failure probability
of~\eqref{eq:coefficient-energy}, proves~\eqref{eq:estimation-tail}.
\end{proof}

The energy event does not depend on $F$, so its failure probability
$|\mathcal T|e^{-R}$ is paid once for the whole family $\mathcal P$, while
each $F\in\mathcal P$ pays only $2e^{-\ell_F}$. This is what makes the
lemma usable for exponentially many projections: in
Section~\ref{sec:est}, the tail parameters~\eqref{eq:tail-parameter} keep
the sum of these terms small by~\eqref{eq:union}, and
Lemma~\ref{lem:class-separation} verifies the sign separation within
each class.

\section{Preserving all cuts}\label{sec:allcuts}

We keep the accurate query cuts fixed, so the only randomness is in the
three tables. A cut of $G$ is the disjoint union of its restrictions to
the connected components of $G$, so it suffices to handle one component
at a time. Fix a connected component of $G$ and let $E$ denote its
positive edges. Cuts $S$ are taken within this component, while $V$ and
$n$ still refer to the whole graph. We split each cut into its
projections onto connectivity classes (Section~\ref{sec:classes}), bound
the error on all projections simultaneously
(Sections~\ref{sec:decompose}--\ref{sec:est}), and then sum over classes
and components (Section~\ref{sec:cuts}). Section~\ref{sec:correctness}
collects the resulting guarantee.

\subsection{Connectivity classes and cut projections}\label{sec:classes}

Partition the positive edges by their connectivity estimates:
\[
 E_\nu=\{e\in E:\ K\le\kappa_e<2K\},\qquad K=2^\nu.
\]
These classes are used only in the proof and are distinct from the
sampling layers $k$. By~\eqref{eq:calibration}, every positive edge
has $.9\le.9\lambda_e\le\kappa_e\le1.1\lambda_e\le1.1(n-1)U$, so it
suffices to take $\nu=-1,\ldots,\lceil\log_2(2n^2U)\rceil$, and there
are at most $J$ classes. Fix one class. For a cut $S$, its projection onto this class is
$F=E_\nu\cap E(S,V\setminus S)$. The error that this class contributes
to the cut, namely the total estimated weight of the selected edges in
$F$ minus the true weight $w(F)$, depends only on $F$. Cuts with the
same projection therefore have the same error from this class, and it
suffices to union bound over the distinct nonempty projections rather
than over all cuts. (An empty projection contributes no error.)

Two weights play different roles:
\begin{equation}\label{eq:mass}
 m_F=w(F),\qquad
 W_F=\min\bigl\{w(S,V\setminus S):\ E_\nu\cap E(S,V\setminus S)=F\bigr\}.
\end{equation}
The projection mass $m_F$ controls the variance. The smallest realizing
cut weight $W_F$ controls how many such projections there can be.
Since $F$ lies in every cut that realizes it, $m_F\le W_F$. Neither quantity is computed by the decoder.

By~\eqref{eq:calibration}, every edge of the class has true
connectivity at least $K/1.1$, and hence every nonempty projection has
$W_F\ge K/1.1$. Lemma~\ref{lem:counting}, applied to the component
(which has at most $n$ vertices) with $\lambda_0=K/1.1$ and
$F^\star=E_\nu$, gives
\[
 \#\{F\ne\varnothing:\ W_F\le\alpha K/1.1\}\le n^{2\alpha}
 \qquad(\alpha\ge1).
\]
Group the nonempty projections by $2^\gamma\le1.1W_F/K<2^{\gamma+1}$,
with $\gamma\ge0$. Taking $\alpha=2^{\gamma+1}$ shows that group $\gamma$
has at most $n^{4\cdot2^\gamma}$ members. Thus the choice
\begin{equation}\label{eq:tail-parameter}
 \ell_F=c_0(W_F/K)\,\ell
\end{equation}
with a sufficiently large absolute constant $c_0$ ensures
\begin{equation}\label{eq:union}
 \sum_{\text{components, classes}}\ \sum_{F\ne\varnothing}e^{-\ell_F}
       \le\delta/1000.
\end{equation}
Indeed, every member of group $\gamma$ has
$\ell_F\ge c_0\ell\,2^\gamma/1.1$, so the contribution of the group is at
most $n^{4\cdot2^\gamma}e^{-c_0\ell2^\gamma/1.1}
=\exp\{-[c_0\ell/1.1-4\ln n]2^\gamma\}$. Moreover, there are at most
$nJ$ component--class pairs. We will use this same union bound for each of
the three errors.

\subsection{Decomposing the error}\label{sec:decompose}

Our goal for every nonempty projection is
\begin{equation}\label{eq:projection-error}
 \Bigl|\sum_{e\in F\cap E_\star}Z_e-m_F\Bigr|
 \le C\Bigl(\sqrt{\frac{m_FW_F\ell}{\beta}}+\frac{W_F\ell}{\beta}\Bigr).
\end{equation}
For the calculations, abbreviate
\[
 x_F=W_F\ell/\beta,\qquad g_F=\sqrt{m_Fx_F}+x_F.
\]
The same scale $g_F$ will bound all three error terms.

Selection success depends on the survival bits at every layer,
including the coins $\xi_{e,k_e+1}$, so we do not condition on it.
Instead, we clip the extra-layer decision:
\[
 \bar\eta_e=\eta_e\,\ind\bigl[w_e2^{k_e}<5t_e(0)/12\bigr].
\]
The indicator is deterministic, so $\bar\eta_e$, like $\eta_e$, is
determined by the selection seeds and the survival bits through layer
$k_e$. Lemma~\ref{lem:twostep} shows that $\bar\eta_e=\eta_e$ on
selection success, since $7/24<5/12$. Unlike $\eta_e$, however,
$\bar\eta_e$ keeps the coefficients of the extra-layer error bounded on
every realization.

Define the three errors:
\begin{align}
 X_1(F)&=\sum_{e\in F}w_e\bigl(2^{k_e}I_{e,k_e}-1\bigr),
       &&\text{baseline sampling},\label{eq:X1}\\
 X_2(F)&=\sum_{e\in F}w_e\,\bar\eta_e\,2^{k_e}I_{e,k_e}
                         \bigl(2\xi_{e,k_e+1}-1\bigr),
       &&\text{one extra layer},\label{eq:X2}\\
 X_3(F)&=\sum_{e\in F\cap E_\star}2^{j_e}\bigl(\widehat w_e-w_e\bigr),
       &&\text{weight estimation}.\label{eq:X3}
\end{align}
On selection success,
\begin{equation}\label{eq:error-split}
 \sum_{e\in F\cap E_\star}Z_e-m_F=X_1(F)+X_2(F)+X_3(F).
\end{equation}
To see this, recall that $Z_e=2^{\widehat j_e}\widehat w_e$. On selection
success, Lemma~\ref{lem:twostep} gives $\widehat j_e=j_e$ for selected
edges and $e\in E_\star\Leftrightarrow I_{e,j_e}=1$. Hence
\[
 \sum_{e\in F\cap E_\star}Z_e-m_F
 =\sum_{e\in F}w_e\bigl(2^{j_e}I_{e,j_e}-1\bigr)+X_3(F).
\]
The first sum splits into $X_1(F)+X_2(F)$ by the identity
\[
 2^{k+\eta}I_{e,k+\eta}-1
 =(2^kI_{e,k}-1)+\eta\,2^kI_{e,k}(2\xi_{e,k+1}-1),
 \qquad \eta\in\{0,1\},
\]
applied with $k=k_e$ and $\eta=\eta_e=\bar\eta_e$. The identity is
trivial for $\eta=0$, and for $\eta=1$ it follows from nested survival,
$I_{e,k+1}=I_{e,k}\,\xi_{e,k+1}$. The first term on the right is the
error of sampling $e$ at its baseline layer. The second is the change
caused by one more fair coin: if $e$ survives to layer $k$, its rescaled
indicator $2^kI_{e,k}$ doubles or drops to zero with equal probability.
Thus the adaptive choice of the layer enters only through the
coefficients $\bar\eta_e$ of $X_2(F)$.

It remains to bound each $X_i(F)$ by a constant multiple of $g_F$.

\subsection{Baseline sampling: Bernstein's inequality}\label{sec:baseline}

The baseline layers $k_e$ are fixed once the query cuts are fixed. Thus
the summands of $X_1(F)$ are independent and centered. Put
$D_\nu=5K/(6\beta)$. If $k_e\ge1$, minimality
in~\eqref{eq:baseline-layer} and $t_e(0)=\kappa_e/\beta<2K/\beta$ give
\[
 w_e2^{k_e}<5t_e(0)/12<D_\nu.
\]
If $k_e=0$, the summand is identically zero. It follows that the
summands have magnitude at most $D_\nu$ and $\Var X_1(F)\le D_\nu m_F$.

By Bernstein's inequality, in the form~\eqref{eq:freedman}, we have
$|X_1(F)|=O(\sqrt{D_\nu m_F\ell_F}+D_\nu\ell_F)$ except with probability
at most $2e^{-\ell_F}$. Since $D_\nu\ell_F=(5c_0/6)x_F$, this bound is
$O(g_F)$. A union bound over all projections, using~\eqref{eq:union},
shows that
\begin{equation}\label{eq:baseline}
 |X_1(F)|\le C_1g_F
\end{equation}
holds simultaneously for all of them, with failure probability at most
$\delta/20$.

\subsection{The extra layer: Freedman's inequality}\label{sec:extra}

Order the edges of $F$ as $e_1,\ldots,e_m$ by increasing
$k_i=k_{e_i}$, with deterministic ties, and write
\[
 X_2(F)=\sum_{i=1}^ma_i\bigl(2\xi_{e_i,k_i+1}-1\bigr),\qquad
 a_i=w_{e_i}\,\bar\eta_{e_i}\,2^{k_i}I_{e_i,k_i}.
\]
To view this sum as a martingale, reveal the randomness in the
following order: first the selection seeds, and then the survival bits
layer by layer, in increasing order of the layer. Within a layer, reveal
the coins $\xi_{e_i,k_i+1}$ of the edges of $F$ in increasing order of
$i$, and the other coins in any order. Since the $k_i$ are
nondecreasing, the coins $\xi_{e_1,k_1+1},\ldots,\xi_{e_m,k_m+1}$ are
revealed in this order. For $1\le i\le m$, let $\mathcal H_{i-1}$
denote everything revealed before $\xi_{e_i,k_i+1}$, and let
$\mathcal H_m$ denote everything revealed. The coefficient $a_i$
depends only on the selection seeds and the survival bits through layer
$k_i$ (Section~\ref{sec:twostep}), so it is determined by
$\mathcal H_{i-1}$. In contrast, $\xi_{e_i,k_i+1}$ is independent of
everything revealed before it, so $2\xi_{e_i,k_i+1}-1$ is a fresh
Rademacher sign given $\mathcal H_{i-1}$. Hence the partial sums of
$X_2(F)$ form a martingale with respect to $(\mathcal H_i)$. The
clipping ensures $0\le a_i\le D_\nu$ on every realization, including
those where a sketch query is inaccurate.
Therefore the predictable quadratic variation $V_2(F)$ of $X_2(F)$, as
in Lemma~\ref{lem:freedman}, satisfies
\begin{equation}\label{eq:correction-variance}
 V_2(F)=\sum_ia_i^2
 \le D_\nu\sum_{e\in F}w_e2^{k_e}I_{e,k_e}
 =D_\nu\bigl(m_F+X_1(F)\bigr).
\end{equation}

The increments $d_i=a_i(2\xi_{e_i,k_i+1}-1)$ are martingale
differences, have conditional variance $a_i^2$, and are bounded by
$D_\nu$ on every realization. Whenever~\eqref{eq:baseline} holds,
$V_2(F)$ is at most the deterministic quantity
$v=D_\nu(m_F+C_1g_F)$. Inequality~\eqref{eq:freedman} bounds the joint event
of this variance bound and a deviation exceeding
$\sqrt{2v\ell_F}+\frac23D_\nu\ell_F$. This threshold is $O(g_F)$: its
linear term is $O(x_F)$, and
\begin{equation}\label{eq:absorb}
 \sqrt{(m_F+C_1g_F)x_F}
 \le\sqrt{m_Fx_F}+\sqrt{C_1g_Fx_F}
 \le(1+\sqrt{C_1})g_F.
\end{equation}
The last inequality uses $x_F\le g_F$. Inequality~\eqref{eq:union} now
gives $|X_2(F)|\le C_2g_F$ for all projections, with an additional
failure probability at most $\delta/20$. We do not condition on~\eqref{eq:baseline}.
Instead, if $|X_2(F)|$ exceeds the threshold, then
either~\eqref{eq:baseline} fails, which is already accounted for in
Section~\ref{sec:baseline}, or the joint event bounded
by~\eqref{eq:freedman} occurs.

\subsection{Weight estimation: separating the signs within a class}
\label{sec:est}

Fix any survival bits, selection seeds, estimation hashes, and
diagnostic signs such that $\mathcal E_{\rm pre}$
and~\eqref{eq:baseline} hold. Then only the
evaluation signs remain random. For a selected edge, define
\[
 u_e=2^{j_e}v_e,\qquad \pi_e=(j_e,r_e).
\]
Lemma~\ref{lem:diagnostic} gives $\norm{u_e}_2\le t_e(0)/18$. Fix one
connectivity class $E_\nu$ and put $\mathcal T=E_\star\cap E_\nu$. Two
targets in this class using the same layer and row cannot appear in one
another's collision vectors: each is too heavy for the other's
collision-norm bound.

\begin{lemma}[Separation within a connectivity class]
\label{lem:class-separation}
For $e,f\in\mathcal T$ with $\pi_e=\pi_f$, we have $u_e(f)=0$.
Consequently, no decoding sign of this class occurs in the collision
noise of any target in this class.
\end{lemma}
\begin{proof}
The case $f=e$ follows from the definition of $v_e$. Otherwise, the two
targets use the same layer $j$ and row. Since $j=j_f\ge k_f$, the
baseline definition implies $w_f\ge5t_f(k_f)/24\ge5t_f(j)/24$. Also
$\kappa_f\ge K>\kappa_e/2$ within the class, so
\[
 w_f\ge\frac5{24}t_f(j)>
          \frac5{48}t_e(j)>
          \frac1{18}t_e(j)\ge\norm{v_e}_2.
\]
A selected $f$ survives at layer $j$. If $v_e(f)\ne0$, its magnitude
would therefore be exactly $w_f$, contradicting this inequality. Thus
$v_e(f)=u_e(f)=0$.

For the second statement, recall that the collision noise of $e$ uses
only the signs $\sigma^{\pi_e,l}$. If $\pi_f\ne\pi_e$, it therefore
contains no decoding sign $\sigma_f^{\pi_f,l}$ of $f$. If $\pi_f=\pi_e$,
the coefficient of $\sigma_f^{\pi_e,l}$ is $u_e(f)=0$ by the first
statement.
\end{proof}

By~\eqref{eq:edgeerror}, for a projection $F$ onto this class,
\[
 X_3(F)=\frac1R\sum_{e\in F\cap\mathcal T}\sum_{l=1}^R
            \sigma_e^{\pi_e,l}\sum_fu_e(f)\,\sigma_f^{\pi_e,l}.
\]
This is the quantity $Q_F$ of Lemma~\ref{lem:estimation}, whose
hypothesis is verified by Lemma~\ref{lem:class-separation}. It remains
to bound the energy of the collision noise.

Every selected edge survives its baseline layer, and
$t_e(0)\le(24/5)w_e2^{k_e}$ by~\eqref{eq:baseline-layer}. Also
$t_e(0)<2K/\beta$ in this class. Consequently,
\begin{equation}\label{eq:est-variance}
 \sum_{e\in F\cap\mathcal T}\norm{u_e}_2^2
 \le C\frac K\beta\sum_{e\in F}w_e2^{k_e}I_{e,k_e}
 \le C\frac K\beta\bigl(m_F+C_1g_F\bigr).
\end{equation}
Use the right side of~\eqref{eq:est-variance} for $v_F$
and~\eqref{eq:tail-parameter} for $\ell_F$ in
Lemma~\ref{lem:estimation}. Its threshold is at most
\[
 C\sqrt{(m_F+C_1g_F)x_F/R}\le C_3g_F.
\]
This controls all projections within the fixed class.

Apply the lemma separately to each component and connectivity class,
then union bound their failure probabilities. The target sets partition
$E_\star$, so the failure terms for their common energy bounds sum to
\[
 \sum_{\text{components, classes}}|\mathcal T|e^{-R}
       =|E_\star|e^{-R}\le Ne^{-R}.
\]
Together with~\eqref{eq:union} and $R=C_R\ell+O(1)$, this bounds the
total estimation failure probability by $\delta/20$.

The conditioning above is performed separately for each class: a decoding
sign from one class may occur in another class's collision noise. The
final union bound needs no independence between classes. Since this
bound holds for every such choice of the other randomness, it also holds
after averaging over these choices.

\subsection{From projections to cuts}\label{sec:cuts}

On selection success, \eqref{eq:error-split} and the bounds of
Sections~\ref{sec:baseline}--\ref{sec:est} prove~\eqref{eq:projection-error}
for all nonempty projections simultaneously. Now consider a cut $S$ of
the component, let $W$ be its weight, and let $F_\nu$ be its nonempty
projections. The classes partition the edges, and $S$ itself realizes
each $F_\nu$, so
\[
 \sum_\nu m_{F_\nu}=W,\qquad W_{F_\nu}\le W.
\]
The error of $H$ on $S$ is the sum of the errors on its projections:
\[
 w_H(S,V\setminus S)-W
 =\sum_\nu\Bigl(\sum_{e\in F_\nu\cap E_\star}Z_e-m_{F_\nu}\Bigr).
\]
Apply~\eqref{eq:projection-error} to each term and use the triangle
inequality. Since there are at most $J$ classes, the Cauchy--Schwarz
inequality bounds the square-root terms by
\[
 \sum_\nu\sqrt{\frac{m_{F_\nu}W_{F_\nu}\ell}{\beta}}
 \le\sqrt{\frac{W\ell}{\beta}}\sum_\nu\sqrt{m_{F_\nu}}
 \le\sqrt{\frac{W\ell}{\beta}}\sqrt{J\sum_\nu m_{F_\nu}}
 =W\sqrt{\frac{J\ell}{\beta}}.
\]
Each linear term is at most $W\ell/\beta$, so these terms sum to at most
$JW\ell/\beta$. Hence
\[
 |w_H(S,V\setminus S)-W|
 \le CW\Bigl(\sqrt{\frac{J\ell}{\beta}}+\frac{J\ell}{\beta}\Bigr).
\]
Finally, $\beta\ge C_\beta J\ell/\eps^2$ by~\eqref{eq:params}, so
$\sqrt{J\ell/\beta}\le\eps/\sqrt{C_\beta}$ and
$J\ell/\beta\le\eps^2/C_\beta\le\eps/C_\beta$. For large enough
$C_\beta$, the right-hand side is therefore at most $\eps W$. A cut of $G$ is the disjoint union of its
restrictions to the components of $G$, so summing over the components
gives the same guarantee for every cut of $G$. In particular, cuts of
weight zero are preserved exactly.

\paragraph{Why the masses matter.}
The Cauchy--Schwarz step uses $\sum_\nu m_{F_\nu}=W$ and therefore loses
only a factor $\sqrt J$. If each mass $m_{F_\nu}$ were replaced by its
upper bound $W$, the square-root terms would sum to $JW\sqrt{\ell/\beta}$,
which would require $\beta$ of order $J^2\ell/\eps^2$. Keeping the masses
requires only $\beta$ of order $J\ell/\eps^2$. This saves one
logarithmic factor in the sampling budget, the bucket width, and the
output size. The dependence on $\eps$ remains quadratic.

\paragraph{Output size.}
Finally, we bound the number of selected edges. The baseline layers are
deterministic,
so the indicators $I_{e,k_e}$ are independent over all positive edges
$e$ of $G$, in all components. By the definition of $k_e$,
$\Pp(I_{e,k_e}=1)=2^{-k_e}\le\min\{1,24\beta w_e/(5\kappa_e)\}$. Hence
\[
 \E\sum_{e:\,w_e>0}I_{e,k_e}
 \le\sum_{e:\,w_e>0}\min\{1,24\beta w_e/(5\kappa_e)\}
 \le(16/3)\beta(n-1),
\]
where the last inequality uses $\kappa_e\ge.9\lambda_e$
from~\eqref{eq:calibration} and Lemma~\ref{lem:sum-connectivity}. By
Lemma~\ref{lem:chernoff},
\begin{equation}\label{eq:output-size}
 \sum_{e:\,w_e>0}I_{e,k_e}\le8n\beta
\end{equation}
with failure probability at most $\delta/20$. On selection success,
Lemma~\ref{lem:twostep} and $j_e\ge k_e$ give
\[
 |E_\star|=\sum_{e:\,w_e>0}I_{e,j_e}\le\sum_{e:\,w_e>0}I_{e,k_e}.
\]
Thus~\eqref{eq:output-size} implies $|E_\star|\le8n\beta$. In particular,
the list of selected pairs stays within the limit $\lceil8n\beta\rceil$
of Section~\ref{sec:recovery}.

\subsection{Correctness for accurate query cuts}\label{sec:correctness}

The following proposition collects the results of
Sections~\ref{sec:local}--\ref{sec:allcuts}.

\begin{proposition}[Correctness for accurate query cuts]
\label{prop:correctness}
Fix accurate query cuts independent of the three tables, and let the
survival bits and evaluation signs be fully independent. Outside eight
events, each of probability at most $\delta/20$, the decoder of
Section~\ref{sec:recovery} returns a positive reweighting $H$ of a set
$E_\star\subseteq E$ of at most $8n\beta=O(n\eps^{-2}\log^2n)$ edges,
and $H$ is a $(1\pm\eps)$-cut sparsifier of $G$.
\end{proposition}
\begin{proof}
The eight events are the load bound and selection accuracy
(Lemma~\ref{lem:selection}), the diagnostic
(Lemma~\ref{lem:diagnostic}), positivity of the estimates
(Lemma~\ref{lem:estimate}), the baseline, extra-layer, and
weight-estimation bounds (Sections~\ref{sec:baseline}--\ref{sec:est}),
and the output-size bound~\eqref{eq:output-size}. Outside these events,
Lemma~\ref{lem:twostep} shows that the selected pairs are positive
edges,~\eqref{eq:output-size} gives $|E_\star|\le8n\beta$, and
Lemma~\ref{lem:estimate} shows that all output weights are positive.
Hence the decoder returns $H$. Section~\ref{sec:cuts} shows that every
cut of $G$ is preserved within a factor of $1\pm\eps$.
\end{proof}

With the coarse sparsifier of Theorem~\ref{thm:coarse} and exact
Gomory--Hu trees, the query cuts are accurate whenever the coarse sketch
succeeds, and they
are independent of the three tables because the coarse sketch uses
independent randomness. Proposition~\ref{prop:correctness} therefore
gives a polynomial-time algorithm that, with fully independent survival
bits and evaluation signs, fails with probability at most $9\delta/20$.
Section~\ref{sec:nearlinear} implements this algorithm in near-linear
time and with short seeds.

\section{Near-linear-time implementation}\label{sec:nearlinear}

To achieve the bounds of Theorem~\ref{thm:main}, we modify the algorithm
of Sections~\ref{sec:algorithm}--\ref{sec:allcuts} in three places.
First, the decoder tests all $\Theta(n^2)$ vertex pairs. Second, the
exact Gomory--Hu trees require $n-1$ maximum-flow computations. Third,
the subsampling coins and evaluation signs, which the analysis treats as
fully independent, must be generated from short seeds, quickly enough
for every update and every recovery query.
Sections~\ref{sec:discovery}, \ref{sec:query-forest},
and~\ref{sec:tapes} address these three points in turn.
Section~\ref{sec:coarse} uses the same tools to prove
Theorem~\ref{thm:coarse}, and Section~\ref{sec:proof-main} proves
Theorem~\ref{thm:main}.

\subsection{Fast candidate discovery}\label{sec:discovery}

The decoder of Section~\ref{sec:recovery} tests every eligible pair,
which takes $\Theta(n^2)$ queries even though at most
$\lceil8n\beta\rceil$ pairs are selected. We replace pair enumeration by
a discovery step that reads candidate pairs directly off the selection
buckets. The idea is simple. If one pair dominates a bucket, the bucket
has the sign of that pair's contribution. If we also keep counters in
which this contribution is multiplied by $\pm1$ according to one bit of
the pair's identifier, then comparing their signs with the sign of the
base counter reveals the identifier bit by bit.

\paragraph{Identifiers and additional counters.}
We identify the pairs with $\{0,\ldots,N-1\}$ through the bijection
\[
 \id(\{u,v\})=\frac{(v-1)(v-2)}2+u-1\qquad(1\le u<v\le n).
\]
Given $m$, the pair $\id^{-1}(m)$ is found in $O(\log n)$ time by binary
search for the largest $v$ with $(v-1)(v-2)/2\le m$. Identifiers are
written with $d=\lceil\log_2N\rceil$ bits, and $\bit_j(m)$ denotes the
$j$th bit of $m$. Next to the base selection counter $\Sel$, every
selection bucket stores one bit counter for each $j\in\{0,\ldots,d-1\}$:
\begin{equation}\label{eq:bit-counter}
 \Sel_j^{k,r}(v,b)=
 \sum_{e:\,h_S^{k,r}(e)=b}\chi_v(e)\,I_{e,k}\,s_e^{k,r}\,
                   (-1)^{\bit_j(\id(e))}.
\end{equation}
A bit counter uses the same hash, survival indicators, and selection
signs as the base counter. It only multiplies the contribution of each
pair by a fixed sign determined by the pair's identifier. To leave room
for the $d$ additional counters in the union bound below, we enlarge the
selection width from $B$ to $B_S$, the smallest power of two at least
$C_B\beta(d+1)$, and let the selection hashes map to $[B_S]$. The
diagnostic and estimation tables are unchanged. The enlarged selection
table still satisfies Lemma~\ref{lem:selection}, whose proof needs only
a width of at least $C_B\beta$.

\paragraph{Decoding candidates.}
After the stream, we compute the cut sums of all counters on every
fundamental cut $S_q$, by the postorder aggregation described at the end
of this section. For every fundamental cut $S_q$, layer $k$, row $r$,
and bucket $b$ with $\Sel^{k,r}(S_q,b)\ne0$, we decode
\[
 \widehat m=
 \sum_{j=0}^{d-1}
 \ind\bigl[\Sel_j^{k,r}(S_q,b)\,\Sel^{k,r}(S_q,b)<0\bigr]\,2^j.
\]
In words, bit $j$ is set exactly when the bit counter and the base
counter have opposite signs. We discard values $\widehat m\ge N$ and
pairs $\id^{-1}(\widehat m)$ that are not eligible, and we remove
duplicates by sorting. The remaining pairs are the \emph{candidates}.
For each candidate $e$ we run the selection rule~\eqref{eq:selection} at
its own query cut $S_e$, followed by the diagnostic and the estimate,
exactly as in Section~\ref{sec:recovery}. A decoded identifier is
therefore only a candidate: whether it enters $E_\star$ is decided by
the same test as in pair enumeration.

\begin{proposition}[Candidate discovery]\label{prop:discovery}
Fix accurate query cuts and any survival bits such that the load
bound~\eqref{eq:load} holds. With failure probability at most
$\delta/20$ over the selection seeds, every positive edge $e$ with
$I_{e,k_e}=1$ is a candidate. If, in addition,~\eqref{eq:pointgood}
holds, then candidate discovery and pair enumeration return the same
output.
\end{proposition}
\begin{proof}
\emph{A dominated bucket decodes correctly.} Fix a positive edge $e$
with $I_{e,k_e}=1$ and a row $r$, and write $k=k_e$, $t=t_e(k)$, and
$b=h_S^{k,r}(e)$. The query cut $S_e$ is a fundamental cut, so its
buckets are decoded. In the cut $S_e$, the base counter of bucket $b$
equals $c+\zeta_0$, and the $j$th bit counter equals
$(-1)^{\bit_j(\id(e))}c+\zeta_{j+1}$. Here $c=\sgn_e\,w_e\,s_e^{k,r}$ is
the contribution of $e$, and $\zeta_0,\ldots,\zeta_d$ collect the
contributions of the other pairs in the bucket. By the definition of
$k_e$, $|c|=w_e\ge5t/24$. If $|\zeta_j|\le t/24$ for every $j$, then
$|c|$ exceeds every $|\zeta_j|$, so each counter is nonzero and has the
sign of its term involving $c$. The product of the $j$th bit counter and
the base counter is therefore negative exactly when bit $j$ of $\id(e)$
is one, and the bucket decodes to $\id(e)$.

\emph{A good row exists.} Call row $r$ good for $e$ if $e$ collides with
no other surviving pair of weight above $t$ that crosses $S_e$, and if
the remaining collision noise in each of the $d+1$ counters has
magnitude at most $t/24$. As in the proof of
Lemma~\ref{lem:selection}, the load bound gives at most $2.5\beta$ such
heavy survivors and light squared mass at most $2.5\beta t^2$. A heavy
collision occurs with probability at most $2.5\beta/B_S$. Multiplying
the contribution of each pair by a fixed sign does not change the second
moment of the light noise, so by Chebyshev's inequality each of the
$d+1$ counters has light noise above $t/24$ with probability at most
$1440\beta/B_S$. By a union bound, the row is good with probability at
least $1-(2.5+1440(d+1))\beta/B_S\ge.95$ for a large enough $C_B$, since
$B_S\ge C_B\beta(d+1)$. Rows are independent, so all of them are bad
with probability at most $.05^R$. A union bound over the $N$ pairs
bounds the total failure probability by $N\cdot.05^R\le\delta/20$.

\emph{Same output as enumeration.} Assume that both events hold. By
Lemma~\ref{lem:twostep}, pair enumeration selects exactly the positive
edges $e$ with $I_{e,j_e}=1$, at layer $j_e$. Since $j_e\ge k_e$, each
such edge also has $I_{e,k_e}=1$ and is therefore a candidate, even if
its selection query first passes at layer $k_e+1$. Conversely, every
candidate undergoes the same selection test as in enumeration, so it is
selected exactly when enumeration selects it. In particular, nonedges
are rejected as before. The two decoders therefore select the same pairs
at the same layers, and their diagnostics and estimates coincide.
\end{proof}

The proposition is the only new ingredient for correctness: the
selection and weight-estimation analyses apply unchanged.

\paragraph{Recovery time.}
Candidate discovery removes the enumeration of pairs, but the decoder
must still form the counters of all query cuts and locate the query cut
of every candidate. A naive implementation of either step takes
quadratic time. We
now show that all steps of recovery from the three tables take
near-linear time. Suppose that
the query cuts are induced through~\eqref{eq:query-cut} by a rooted
forest on $V$, as for exact Gomory--Hu trees and for the query forests
of Section~\ref{sec:query-forest}. The selection table now has $d+1$
counters per bucket and $B_S$ buckets per row, so the three tables have
\[
 N_{\rm ctr}=n(J+1)\bigl(RB_S(d+1)+65R^2B\bigr)
 =O\bigl(nJ\beta R((d+1)^2+R)\bigr)=\widetilde O(n/\eps^2)
\]
counters in total. Recovery from the tables consists of the following
steps.
\begin{itemize}
\item \emph{Preprocessing the forest.} A depth-first traversal of each
tree records entry and exit times. For the tree edge $q$ between a
vertex $x$ and its parent, a vertex of the same tree lies in $S_q$
exactly when its entry time lies outside the interval of $x$, which
takes $O(1)$ time to test. Binary lifting stores, for every vertex $x$
and every $h\le\log_2n$, the $2^h$th ancestor of $x$ and a
minimum-weight edge among the $2^h$ tree edges above $x$. After
$O(n\log n)$ preprocessing, a minimum-weight edge $q(e)$ on the tree
path of a pair $e$ is found in $O(\log n)$ time.
\item \emph{Aggregation.} In each tree, we replace every vertex counter,
in place and in postorder, by the sum of the counters in its subtree.
For the tree edge $q$ above $x$, the counters of $S_q$ are then the
differences between the aggregated counters of the root of the tree and
those of $x$. This takes $O(N_{\rm ctr})$ operations.
\item \emph{Discovery.} The at most $n-1$ fundamental cuts have
$(n-1)(J+1)RB_S$ selection buckets in total. Each nonzero bucket is
decoded in $O(d)$ operations, and its pair is recovered in $O(\log n)$
time. The raw list therefore has
$N_{\rm id}\le(n-1)(J+1)RB_S=\widetilde O(n/\eps^2)$ entries, and
sorting it takes $O(N_{\rm id}\log N_{\rm id})$ operations.
\item \emph{Testing candidates.} A candidate needs $q(e)$ and $\sgn_e$,
and then $(J+1)R$ selection queries, each of which reads one aggregated
counter and evaluates one hash value and one sign. Including the
medians, this takes $O(N_{\rm id}(JR+\log n))$ operations and
coefficient evaluations.
\item \emph{Diagnostics and estimation.} Each of the at most
$\lceil8n\beta\rceil$ selected edges needs $64R^2$ diagnostic and $R$
estimation counters with their signs, for $O(n\beta R^2)$ operations
and coefficient evaluations in total.
\end{itemize}
Altogether, recovery from the tables takes $\widetilde O(n/\eps^2)$
operations and coefficient evaluations. Section~\ref{sec:tapes} and
Appendix~\ref{app:precision} compute every coefficient in
polylogarithmic time, so this part of recovery takes
$\widetilde O(n/\eps^2)$ time. By contrast, pair enumeration spends
$\Theta(n^2JR)$ operations on selection queries alone.

\subsection{Query forests from approximate Gomory--Hu trees}
\label{sec:query-forest}

Exact Gomory--Hu trees give more than the analysis needs.
By~\eqref{eq:calibration}, a query cut only has to separate the
endpoints of its pair and have weight within a constant factor of their
connectivity, provided that the query weight is the exact weight of the
cut in the coarse graph. Approximate Gomory--Hu trees
(Theorem~\ref{thm:approx-gh}) provide such cuts in near-linear time: for
every pair, a minimum-weight edge on its tree path has a fundamental cut
of nearly minimum weight.

The difficulty is that Theorem~\ref{thm:approx-gh} needs a connected
graph, and its failure probability is polynomially small only in the
number of vertices of its input, whereas the coarse graph may have many
small components. Running the algorithm once on each component
therefore does not give a small failure probability. One way around
this is to boost: run the algorithm $O(\log n)$ times on each component,
and for each pair use the run that gives the smallest query
weight. Every tree
edge on the path between two vertices has a fundamental cut that
separates them, so query weights never underestimate connectivities. Hence one successful run per component suffices, and each run succeeds with probability at
least $1/2$. The query cuts then come from $O(\log n)$ forests, which costs
an extra logarithmic factor in the recovery time.

Below we give a construction that needs only one call and yields a
single forest, which saves this logarithmic factor. Specifically, we add
a hub vertex and join it to every component of the coarse graph by an edge
much lighter than any cut inside a component. The resulting padded graph
is connected and has $n+1$ vertices, so it satisfies the hypotheses of
Theorem~\ref{thm:approx-gh}, and a single call succeeds with high
probability. The approximate tree of the padded graph may join different
components. However, every tree path between different components
contains an edge whose fundamental cut is light, whereas no tree path
inside a component does. Deleting these edges therefore separates the
components again and leaves a tree on each of them.

\begin{lemma}[Query forest]\label{lem:query-forest}
Let $H$ be a graph on $V$ with $m_H$ edges, whose weights are positive
multiples of $2^{-\mu}$ and at most $n^{O(1)}$, where
$2^\mu\le n^{O(1)}$, and let $\delta_1>0$ with
$\delta_1^{-1}\le n^{O(1)}$. Let $C_1,\ldots,C_t$ be the connected
components of $H$ with at least two vertices. There is a randomized
algorithm that uses $\widetilde O(m_H+n)$ time and space and, with
probability at least $1-\delta_1$, returns for each $a\in[t]$ a tree
$T_a$ on $C_a$, rooted at $r_a=\min C_a$, with edge weights $\tau_q$ such
that the following hold, where the fundamental cut $S_q$ of a tree edge
$q$ is the side of $T_a-q$ that contains $r_a$.
\begin{enumerate}
\item $\tau_q=w_H(S_q,V\setminus S_q)>0$ for every tree edge $q$.
\item $\lambda_H(u,v)\le\tau_q\le1.01\,\lambda_H(u,v)$ for all distinct
$u,v\in C_a$ and every minimum-weight edge $q$ on the $u$--$v$ path of
$T_a$.
\end{enumerate}
\end{lemma}
In other words, $T_a$ is a $1.01$-approximate Gomory--Hu tree of $C_a$ in
which the weight of every tree edge is the exact weight of its
fundamental cut.
\begin{proof}
Find the components of $H$ by breadth-first search, and return the empty
forest if $t=0$.

\emph{Fundamental cut weights.} We use the following subroutine twice.
Given a rooted tree $T$ and a weighted graph $F$ on the same $k$
vertices, it computes, for every tree edge $q$, the weight in $F$ of the
fundamental cut of $q$, that is, the total weight of the edges of $F$
with exactly one endpoint in the subtree below $q$. We keep a number
$\alpha(v)$ at every vertex $v$, initially zero. The idea is to record
the weight of each edge at its two endpoints and to cancel both copies
at their lowest common ancestor, so that the sum of $\alpha$ over a
subtree retains the weight of exactly those edges that leave the
subtree. Concretely, for every edge $\{x,y\}$ of $F$, add its weight to
$\alpha(x)$ and to $\alpha(y)$, and subtract twice its weight from
$\alpha(z)$, where $z$ is the lowest common ancestor of $x$ and $y$. A
subtree contains $z$ exactly when it contains both $x$ and $y$, so the
sum of $\alpha$ over a subtree counts the weight of $\{x,y\}$ exactly
when one of its endpoints lies in the subtree and the other does not.
The desired weights are therefore the subtree sums of $\alpha$, which
one postorder traversal computes. Lowest common ancestors take $O(1)$
time after $O(k\log k)$ preprocessing with an Euler tour and a sparse
table, so the subroutine takes $O(|E(F)|+k\log k)$ time.

\emph{The padded graph.} Multiply every weight of $H$ by $2n2^\mu$, so
that every weight becomes an integer between $2n$ and $n^{O(1)}$. Add a hub vertex
$r_\star$, and join it by an edge of weight one to the smallest vertex of
every connected component of $H$, including the isolated vertices. In
particular, $r_\star$ is joined to every root $r_a$. Let $\widehat H$ be
the resulting graph on $\widehat V=V\cup\{r_\star\}$, with weights
$\widehat w$ and pair connectivities $\widehat\lambda$. It is connected
and has $n+1$ vertices, $O(m_H+n)$ edges, and integer weights between
$1$ and $n^{O(1)}$.

\emph{Connectivities in $\widehat H$.} Inside a component, the padding
only rescales connectivities, while different components are joined
only through light edges. More precisely, let $u,v\in C_a$. A cut of
$\widehat H$ that separates $u$ and $v$ restricts to a cut of $C_a$ that
separates them, and the edges of $C_a$ crossing it already have weight
at least $2n2^\mu\lambda_H(u,v)$. Conversely, let $S\subseteq C_a$ be the
side not containing $r_a$ of a minimum cut of $C_a$ separating $u$ and
$v$. Since $r_a$ is the only vertex of $C_a$ with neighbors outside
$C_a$, the cut $(S,\widehat V\setminus S)$ crosses only edges of $C_a$.
Hence $\widehat\lambda(u,v)=2n2^\mu\lambda_H(u,v)\ge2n$, where the last
step uses $\lambda_H(u,v)\ge2^{-\mu}$. If instead $x\in C_a$ and
$y\in\widehat V\setminus C_a$, then $\widehat\lambda(x,y)\le1$, because
the only edge leaving $C_a$ is the hub edge at $r_a$.

\emph{The approximate tree.} Run the algorithm of
Theorem~\ref{thm:approx-gh} on $\widehat H$ with $\eps_0=1/100$. Since
$|\widehat V|=n+1$ and $\delta_1^{-1}\le n^{O(1)}$, a suitable constant
in its high-probability guarantee makes its failure probability at most
$\delta_1$. We stop the algorithm and abort if it exceeds its time
bound, and we also abort if its output is not a spanning tree of
$\widehat V$. Let $\widehat T$ be the tree, rooted arbitrarily, and let
$\widehat\tau_q$ be the weight in $\widehat H$ of the fundamental cut of
a tree edge $q$, computed by the subroutine. On success, the definition
of an approximate Gomory--Hu tree gives, for all $x,y\in\widehat V$, an
edge $q$ on the $x$--$y$ path of $\widehat T$ with
$\widehat\tau_q\le1.01\,\widehat\lambda(x,y)$.

\emph{Splitting the tree.} Keep the tree edges with
$\widehat\tau_q\ge2n$. For $u,v\in C_a$, every edge on the $u$--$v$ path
of $\widehat T$ separates $u$ and $v$, so it has
$\widehat\tau_q\ge\widehat\lambda(u,v)\ge2n$, and the whole path is
kept. For $x\in C_a$ and $y\notin C_a$, on success some edge on the
$x$--$y$ path has $\widehat\tau_q\le1.01\,\widehat\lambda(x,y)\le1.01<2n$,
and this edge is deleted. Hence the component $T_a$ of the kept forest
that contains $r_a$ has vertex set exactly $C_a$, and for $u,v\in C_a$
the $u$--$v$ path of $T_a$ is the $u$--$v$ path of $\widehat T$. We check
that the components of $r_1,\ldots,r_t$ have vertex sets
$C_1,\ldots,C_t$, and abort otherwise.

\emph{Exact weights in $H$.} Root $T_a$ at $r_a$, and compute
$\tau_q=w_H(S_q,V\setminus S_q)$ for every tree edge $q$ by the
subroutine, applied to $T_a$ and the edges of $H$ inside $C_a$. No edge
of $H$ leaves $C_a$, so this counts every edge crossing $S_q$. Moreover,
$\tau_q>0$, because $S_q$ is a nonempty proper subset of $C_a$ and $C_a$
is connected in $H$. This gives property~1.

For property~2, let $u,v\in C_a$. Every edge $q$ on the $u$--$v$ path of
$T_a$ separates $u$ and $v$, so $\tau_q\ge\lambda_H(u,v)$. Conversely,
choose an edge $q^\star$ on this path with
$\widehat\tau_{q^\star}\le1.01\,\widehat\lambda(u,v)
=1.01\cdot2n2^\mu\lambda_H(u,v)$. Deleting $q^\star$ from $T_a$ splits
$C_a$ into its intersections with the two sides of $\widehat T-q^\star$.
Thus $2n2^\mu\tau_{q^\star}$ is the weight in $\widehat H$ of the edges
inside $C_a$ that cross the fundamental cut of $q^\star$, which is at
most $\widehat\tau_{q^\star}$. A minimum-weight edge of the path
therefore has weight at most $\tau_{q^\star}\le1.01\,\lambda_H(u,v)$.

\emph{Time and space.} Breadth-first search and padding take
$O(m_H+n)$ time, and the two uses of the subroutine take
$O(m_H+n\log n)$ time in total. Theorem~\ref{thm:approx-gh} takes
$\widetilde O(m_H+n)$ time because $\eps_0$ is a constant. The space is
bounded by the time.
\end{proof}

\paragraph{Query cuts from the query forest.}
Lemma~\ref{lem:query-forest} needs weights that are positive multiples
of $2^{-\mu}$ and polynomially bounded. We may assume that $H_0$ has
this form, with $2^\mu=\Theta(n^2)$ and weights at most $2nU$, since
rounding $H_0$ as in Appendix~\ref{sec:normalize} enforces it on every
run. Moreover, by Lemma~\ref{lem:rounding}, if $H_0$
satisfies~\eqref{eq:coarse-accuracy} before rounding, then the rounded
graph, which we still call $H_0$, has the same components as $G$ and
satisfies
\begin{equation}\label{eq:coarse}
 .95\,w(S,V\setminus S)\le w_{H_0}(S,V\setminus S)
                 \le1.04\,w(S,V\setminus S)\qquad(S\subseteq V).
\end{equation}
We apply Lemma~\ref{lem:query-forest} to $H_0$ with $\delta_1=\delta/20$
and with randomness independent of the three tables, and we use the
resulting trees and weights in place of the Gomory--Hu trees
in~\eqref{eq:query-cut}. As before, a pair is eligible if its endpoints
lie in a common nontrivial component of $H_0$.

Now suppose that both Theorem~\ref{thm:coarse} and
Lemma~\ref{lem:query-forest} succeed, and let $e$ be an eligible pair.
We check the three conditions of~\eqref{eq:calibration}.
By~\eqref{eq:coarse}, $.95\lambda_e\le\lambda_{H_0}(e)\le1.04\lambda_e$,
and so property~2 gives
\[
 .95\lambda_e\le\lambda_{H_0}(e)\le\kappa_e\le1.01\cdot1.04\,\lambda_e
 <1.1\lambda_e.
\]
The cut $S_e$ separates the endpoints of $e$ because $q(e)$ lies on
their tree path. Finally, by property~1 and~\eqref{eq:coarse},
$w(S_e,V\setminus S_e)\le w_{H_0}(S_e,V\setminus S_e)/.95=\kappa_e/.95$.
Since $H_0$ and $G$ have the same components, the eligible pairs are
exactly those with $\lambda_e>0$, so the query cuts are accurate. They
are also independent of the three tables, as they depend only on the
coarse sketch and on the independent randomness of
Lemma~\ref{lem:query-forest}.

\subsection{A near-linear-time constant-accuracy sparsifier}
\label{sec:coarse}

In this section, we prove Theorem~\ref{thm:coarse}. For constant $\eps$,
our recovery procedure returns a constant-accuracy sparsifier in
near-linear time, but only if it is given accurate query cuts. The
difficulty is that, in the main algorithm, these query cuts come from a
constant-accuracy sparsifier, which is exactly what we are trying to
build. We escape this circularity using the fact that query cuts need to
be correct only up to constant factors (see~\eqref{eq:calibration}). Let
$x$ and $x'$ be weight vectors on the pairs, viewed as graphs. If every
cut of $x'$ is within a factor close to one of the corresponding cut of $x$,
then a near-minimum cut separating two
vertices in $x$ is also a near-minimum cut separating them in $x'$.
Hence query cuts computed from a sparsifier of $x$ are accurate for
$x'$.

Our construction therefore starts from a graph whose minimum cuts are
explicit and moves toward $G$ in small steps, so that each step can take
its query cuts from the sparsifier recovered in the previous one.
Specifically, let $w_{\rm star}\in\{0,1\}^N$ be the indicator vector of
the spanning star with center $1$, whose edges are $\{1,u\}$ for
$2\le u\le n$. Set
\[
 A=8192n^4,\qquad m_0=128nAU,\qquad
 m_{i+1}=\max\{128,\lceil64m_i/65\rceil\},
\]
and let $L$ be the first index with $m_L=128$. We follow the
regularization path
\[
 W_i=Aw+m_iw_{\rm star}\qquad(0\le i\le L).
\]
In $W_0$, the star is so heavy that it dominates every cut, and its
minimum cuts are explicit. Each step lowers the multiple $m_i$ of the
star by a factor of about $65/64$, so consecutive graphs differ by at
most this factor in every coordinate. As a result, the sparsifier
recovered for one graph supplies accurate query cuts for the next. Each
stage runs our recovery procedure in the form of
Lemma~\ref{lem:query-recovery} below. At the end of the path, the star
is negligible on every cut of positive weight in $Aw$, so we remove it
and divide by $A$, which leaves a constant-factor cut approximation of
$G$. A final recovery on $w$ itself then produces a sparsifier supported
on edges of $G$. Each stage uses its own independent copy of the
three-table sketch, and since the sketches are linear, all copies are
maintained in the same single pass.

\paragraph{Outline.}
After the stream, the construction performs the following steps.
\begin{enumerate}
\item \emph{Stage $0$.} Run our recovery procedure
(Lemma~\ref{lem:query-recovery}) on the sketch of $W_0$, with the star
as the query forest. This gives a sparsifier $Y_0$ of $W_0$.
\item \emph{Stages $1,\ldots,L$.} Run our recovery procedure on the
sketch of $W_i$, with query cuts from a query forest of $Y_{i-1}$. This
gives a sparsifier $Y_i$ of $W_i$.
\item \emph{Separating the components.} Using a query forest of $Y_L$,
discard the pairs of $Y_L$ that join different components of $G$, and
divide the remaining weights by $A$. This gives a graph $\Gamma$ that
approximates every cut of $G$ within a constant factor.
\item \emph{The final recovery.} Run our recovery procedure on the sketch
of $w$, with query cuts from a query forest of $\Gamma$. This gives
$H_0$.
\end{enumerate}

\paragraph{Recovery with given query cuts.}
The analysis in Sections~\ref{sec:local}--\ref{sec:allcuts} does not use
the fact that the query cuts come from the coarse sparsifier. It uses
only that they are accurate (Definition~\ref{def:accurate}) and
independent of the three tables. The next lemma records this
observation for an arbitrary integer weight vector $x$ in place of $w$.
For query cuts induced by a rooted forest through~\eqref{eq:query-cut},
a pair is eligible if its endpoints lie in a common tree, and the query
cuts are accurate for $x$ if Definition~\ref{def:accurate} holds with
$x$ in place of $w$.

\begin{lemma}[Recovery with given query cuts]\label{lem:query-recovery}
Let $x\in\{0,1,\ldots,U'\}^N$ and $0<\delta'<1/4$, with
$U',\delta'^{-1}\le n^{O(1)}$. Consider the three-table sketch of $x$,
built with $(U',1/64,\delta')$ in place of $(U,\eps,\delta)$ and with the
random tapes of Section~\ref{sec:tapes}, and its recovery with candidate
discovery. Suppose that the query cuts are induced by a rooted forest
with $O(\log n)$-bit weights, are accurate for $x$, and are independent
of the sketch. Then the sketch uses $\widetilde O(n)$ bits, and
recovery takes $\widetilde O(n)$ time and returns $\widetilde O(n)$
edges. With probability at least $1-\delta'$,
the output is a positive reweighting of pairs in $\{e:x_e>0\}$ that is a
$(1\pm1/64)$-cut sparsifier of $x$.
\end{lemma}
\begin{proof}
Sections~\ref{sec:local}--\ref{sec:allcuts} use the query cuts only
through their accuracy and their independence from the three tables.
Otherwise, they use only that the input weights are integers in
$\{0,\ldots,U\}$: this makes positive weights at least one, gives
$t_e(0)\le1.1(n-1)U$, and lets us apply Lemmas~\ref{lem:sum-connectivity}
and~\ref{lem:counting}. All of this holds for $x$ with $U'$ in place of
$U$. With fully independent tapes,
Propositions~\ref{prop:correctness} and~\ref{prop:discovery}, applied
with the parameters $(U',1/64,\delta')$, therefore bound the failure
probability by $9\delta'/20$. Implementing the tapes as in
Section~\ref{sec:tapes} adds at most $2\delta'/20$, so the total is at
most $11\delta'/20<\delta'$.

The remaining bounds hold on every run. The decoder returns at most
$\lceil8n\beta'\rceil$ pairs (Section~\ref{sec:recovery}), where the
sampling budget $\beta'$ is $O(\log^2n)$ at accuracy $1/64$. The tables
and seeds take $\widetilde O(n)$ bits, and recovery takes
$\widetilde O(n)$ time, by Sections~\ref{sec:discovery}
and~\ref{sec:tapes}.
\end{proof}

\paragraph{Properties of the path.}
A nonempty set $S$ with $1\notin S$ has $w_{\rm star}(S,V\setminus S)=|S|$
and
\[
 w(S,V\setminus S)\le|S|\,(n-1)\,U<nU\,w_{\rm star}(S,V\setminus S).
\]
We use a star because its cut weight $|S|$ grows at the same rate as
this bound on the cuts of $w$. Hence $m_0=128nAU$, only a factor $128n$
larger than the largest weight of $Aw$, already makes the star dominate
every cut of $W_0$. Scaling $w$ by $A$, rather than lowering the multiple
of the star below one, keeps every $W_i$ integral: each $W_i$ is an
integer vector with all coordinates in $\{0,\ldots,U'\}$, where
$U'=AU+m_0$. Its support contains the star, so $W_i$ is connected and
all pairs have positive connectivity in it.

The schedule satisfies
\begin{equation}\label{eq:coarse-schedule}
 W_{i+1}\le W_i\le\frac{65}{64}W_{i+1}
 \quad\text{coordinatewise},\qquad W_L=Aw+128\,w_{\rm star},
\end{equation}
and its length is $L=O(\log(nU))$. For~\eqref{eq:coarse-schedule}, it
suffices to show that $m_{i+1}\le m_i\le\frac{65}{64}m_{i+1}$. The
second inequality holds because $m_{i+1}\ge64m_i/65$. For the first,
$m_i\ge128$ gives $\lceil64m_i/65\rceil\le64m_i/65+1\le m_i$, and also
$128\le m_i$. For the length, if $m_i\ge130$, then
\[
 m_{i+1}\le\max\Bigl\{128,\ \frac{64m_i}{65}+1\Bigr\}
 \le\frac{129}{130}\,m_i,
\]
because $m_i/130\ge1$. Hence $m_i$ decreases geometrically until it is
at most $129$, and from $129$ the next value is
$\lceil64\cdot129/65\rceil=128$. Therefore
$L\le\log_{130/129}(m_0/128)+2=O(\log(nU))$.

\paragraph{Copies and the single pass.}
The length $L$ depends only on $n$ and $U$, so all copies can be
allocated before the stream begins. We use $L+2$ independent copies of
the three-table sketch: one for each $W_i$ with $0\le i\le L$, with
weight bound $U'$, and one for $w$, with weight bound $U$. Every copy
has accuracy $1/64$, implements its random tapes as in
Section~\ref{sec:tapes}, and has failure parameter $\delta'$, the
largest power of two at most $\delta_0/[2(L+2)]$. All copies are
maintained during the single pass over the original stream. Let
$\Phi_i$ be the linear map computed by the copy for $W_i$. Once the
randomness of the copy is fixed, $\Phi_i$ is a fixed matrix. During the
stream we maintain $\Phi_iw$, applying each update to this copy exactly
as in Section~\ref{sec:tables}. After the stream, we form
\begin{equation}\label{eq:regularizer-sketch}
 \Phi_iW_i=A\,\Phi_iw+m_i\,\Phi_iw_{\rm star}
\end{equation}
by multiplying every counter of the copy by $A$ and then applying the
$n-1$ updates $(\{1,u\},m_i)$ for $2\le u\le n$. By linearity, the
result is exactly the sketch that the copy would hold after a stream
that produces $W_i$ directly. If any step below aborts, the
construction returns the empty graph.

\paragraph{Stage $0$: query cuts from the star.}
We say that \emph{stage $i$ succeeds} if the copy for $W_i$ returns a
graph $Y_i$ with
\begin{equation}\label{eq:stage-output}
 \frac{63}{64}W_i(S,V\setminus S)\le Y_i(S,V\setminus S)
                         \le\frac{65}{64}W_i(S,V\setminus S)
 \qquad(S\subseteq V).
\end{equation}
For $W_0$, the star dominates every cut. If $S$ is nonempty and
$1\notin S$, then
$Aw(S,V\setminus S)<nAU\,w_{\rm star}(S,V\setminus S)
=(m_0/128)\,w_{\rm star}(S,V\setminus S)$, so
\[
 m_0\,w_{\rm star}(S,V\setminus S)\le W_0(S,V\setminus S)
 \le\frac{129}{128}\,m_0\,w_{\rm star}(S,V\setminus S).
\]
The same bounds hold when $1\in S\ne V$, by passing to the complement.
We use the star itself as the query forest. Root it at $1$, and give
every edge $q_u=\{1,u\}$ the weight $m_0$, so that its fundamental cut
is $S_{q_u}=V\setminus\{u\}$. For a pair $e=\{u,v\}$ with $u<v$, the
minimum-weight edge on its tree path, with ties broken toward the
smaller child, is $q_v$ if $u=1$ and $q_u$ otherwise. In either case
$S_e=V\setminus\{z\}$ for an endpoint $z\ne1$ of $e$, so $S_e$ separates
the endpoints of $e$, $w_{\rm star}(S_e,V\setminus S_e)=1$, and
\[
 W_0(S_e,V\setminus S_e)\le\frac{129}{128}\,m_0
 =\frac{129}{128}\,\kappa_e\le\kappa_e/.9.
\]
Every cut separating $u$ and $v$ leaves one of them on the side without
vertex $1$, so it has $w_{\rm star}$-weight at least one and
$W_0$-weight at least $m_0$. Hence
$m_0\le\lambda_e(W_0)\le W_0(S_e,V\setminus S_e)\le\frac{129}{128}m_0$,
which gives $\frac{128}{129}\lambda_e(W_0)\le\kappa_e\le\lambda_e(W_0)$.
The star spans $V$, so all pairs are eligible, and all of them have
positive connectivity because $W_0$ is connected. These query cuts are
deterministic and accurate for $W_0$, so Lemma~\ref{lem:query-recovery}
shows that stage $0$ succeeds except with probability at most
$\delta'$.

\paragraph{Stage $i$: query cuts from the previous stage.}
The graph $W_i$ is available only through its sketch, whereas $Y_{i-1}$
is an explicit graph with $\widetilde O(n)$ edges whose cuts are within
a constant factor of those of $W_i$. Its near-minimum cuts are therefore
near-minimum cuts of $W_i$, and we use them as query cuts. Formally, let
$1\le i\le L$, and suppose that stage $i-1$ succeeded.
By~\eqref{eq:coarse-schedule}, every cut satisfies
$W_i(S,V\setminus S)\le W_{i-1}(S,V\setminus S)
\le\frac{65}{64}W_i(S,V\setminus S)$. Combined
with~\eqref{eq:stage-output} for $i-1$, this gives
\[
 \frac{63}{64}W_i(S,V\setminus S)
 \le Y_{i-1}(S,V\setminus S)
 \le\Bigl(\frac{65}{64}\Bigr)^2W_i(S,V\setminus S).
\]
Since $63/64>24/25$ and $(65/64)^2<26/25$, Lemma~\ref{lem:rounding}
applies with $x=W_i$, $V'=U'$, and $Y=Y_{i-1}$: the normalized graph
$\widetilde Y_{i-1}=\mathcal R_{U'}(Y_{i-1})$ approximates every cut of
$W_i$ within $[.95,1.04]$, and it is connected because $W_i$ is. We
check that $\widetilde Y_{i-1}$ is connected, aborting otherwise, and
run Lemma~\ref{lem:query-forest} on it with failure parameter $\delta'$
and fresh randomness. On success it returns a single tree on $V$. As at
the end of Section~\ref{sec:query-forest}, with $W_i$ in place of $w$
and $\widetilde Y_{i-1}$ in place of $H_0$, the induced query cuts
satisfy $.95\lambda_e(W_i)\le\kappa_e\le1.0504\lambda_e(W_i)$ and
$W_i(S_e,V\setminus S_e)\le\kappa_e/.95$, so they
satisfy~\eqref{eq:calibration} for $W_i$. All pairs are eligible, and
all have positive connectivity in the connected graph $W_i$, so the
query cuts are accurate for $W_i$. They are a function of $Y_{i-1}$ and
the fresh randomness, and $Y_{i-1}$ depends only on the copies for
$W_0,\ldots,W_{i-1}$. Hence they are independent of the copy for
$W_i$.
Conditional on the outcome of all earlier steps,
Lemma~\ref{lem:query-recovery} shows that stage $i$ succeeds except with
probability at most $\delta'$.

Each stage recovers $W_i$ from its own exact sketch, and the previous
output supplies only query cuts, that is, it only determines where the
recovery looks. Errors therefore do not accumulate along the path: every
successful stage has accuracy $1/64$ with respect to its own $W_i$, no
matter how many stages precede it.

\paragraph{Separating the components.}
After the last stage, $Y=Y_L$ approximates $W_L=Aw+128\,w_{\rm star}$
within $[63/64,65/64]$. The star is now negligible on every cut of
positive weight in $G$: such a cut has $W_L$-weight at least $A$, while
its star part is at most $128(n-1)$. The star still connects different
components of $G$, however, and these artificial connections must be
removed before we divide by $A$. Normalize
$\widetilde Y=\mathcal R_{U'}(Y)$, check that it is connected, and run
Lemma~\ref{lem:query-forest} on it with failure parameter $\delta'$.
Let $\kappa_e$ be the query weight that the resulting tree assigns to a
pair $e$. For every pair $e$ in the support of $Y$, set
\[
 \Gamma_e=\frac{Y_e}{A}\,\ind[\kappa_e\ge A/2],
\]
and set $\Gamma_e=0$ for all other pairs. If stage $L$ succeeded, then
the argument of the previous paragraphs, now with $x=W_L$ and
$Y=Y_L$, gives $.95\lambda_e(W_L)\le\kappa_e\le1.0504\lambda_e(W_L)$
for every pair on success of Lemma~\ref{lem:query-forest}, and in
particular $.9\lambda_e(W_L)\le\kappa_e\le1.1\lambda_e(W_L)$. With these
bounds, the threshold keeps exactly the pairs whose
endpoints lie in the same component of $G$:
\begin{itemize}
\item If $u$ and $v$ lie in the same component of $G$, every cut
separating them has $w$-weight at least one. Hence
$\lambda_e(W_L)\ge A$, and $\kappa_e\ge.9A\ge A/2$.
\item If $u$ and $v$ lie in different components, let $C$ be the
component containing $u$. Then $w(C,V\setminus C)=0$ and
$W_L(C,V\setminus C)=128\,w_{\rm star}(C,V\setminus C)\le128(n-1)$, so
$\kappa_e\le1.1\cdot128(n-1)<A/2$.
\end{itemize}

\paragraph{The discarded weight is small.}
Let $C_1,\ldots,C_p$ be the components of $G$, including isolated
vertices. A discarded pair has its endpoints in two different
components, so it crosses exactly two of the cuts
$(C_a,V\setminus C_a)$, whereas a kept pair crosses none of them. Every
$C_a$ has $w(C_a,V\setminus C_a)=0$, so~\eqref{eq:stage-output} gives
\[
 2\sum_{e\text{ discarded}}Y_e
 \le\sum_{a=1}^pY(C_a,V\setminus C_a)
 \le\frac{65}{64}\sum_{a=1}^p128\,w_{\rm star}(C_a,V\setminus C_a)
 \le\frac{65}{64}\cdot128\cdot2(n-1),
\]
where the last step holds because each of the $n-1$ star edges crosses
at most two of these cuts. Hence the discarded weight is at most
$130(n-1)$.

\paragraph{$\Gamma$ approximates every cut of $G$.}
Let $S\subseteq V$. If $w(S,V\setminus S)=0$, then $S$ is a union of
components of $G$, so every pair crossing $S$ is discarded and
$\Gamma(S,V\setminus S)=0$. Otherwise $w(S,V\setminus S)\ge1$ by
integrality. The kept part of $Y(S,V\setminus S)$ equals
$A\,\Gamma(S,V\setminus S)$, and the discarded part is at most
$130(n-1)$. Using $Y(S,V\setminus S)\ge\frac{63}{64}W_L(S,V\setminus S)
\ge\frac{63}{64}A\,w(S,V\setminus S)$ and $w(S,V\setminus S)\ge1$, we
get
\[
 \Gamma(S,V\setminus S)
 \ge\frac{63}{64}w(S,V\setminus S)-\frac{130(n-1)}{A}
 \ge\Bigl(\frac{63}{64}-\frac{130}{8192n^3}\Bigr)w(S,V\setminus S)
 >\frac{24}{25}w(S,V\setminus S).
\]
In the other direction,
$A\,\Gamma(S,V\setminus S)\le Y(S,V\setminus S)
\le\frac{65}{64}\bigl(A\,w(S,V\setminus S)+128(n-1)\bigr)$, and
$128(n-1)/A<1/(64n^3)\le1/512$, so
\[
 \Gamma(S,V\setminus S)
 \le\frac{65}{64}\Bigl(1+\frac1{512}\Bigr)w(S,V\setminus S)
 <\frac{26}{25}w(S,V\setminus S).
\]
Thus $\Gamma$ approximates every cut of $w$ within $[24/25,26/25]$. It
may still contain pairs that are not edges of $G$, because $Y$ is
supported on arbitrary pairs. The final recovery removes them.

\paragraph{The final recovery.}
By Lemma~\ref{lem:rounding} with $x=w$, $V'=U$, and $Y=\Gamma$, the
normalized graph $\widetilde\Gamma=\mathcal R_U(\Gamma)$ approximates
every cut of $w$ within $[.95,1.04]$ and has the same components as $G$.
We run Lemma~\ref{lem:query-forest} on $\widetilde\Gamma$ with failure
parameter $\delta'$. If it succeeds, the argument at the end of
Section~\ref{sec:query-forest}, with $\widetilde\Gamma$ in place of
$H_0$, shows that the resulting query cuts are accurate for $w$. They
depend only on the copies for $W_0,\ldots,W_L$ and on the randomness of
the query forests, so they are independent of the copy for $w$.
Lemma~\ref{lem:query-recovery}, applied to this copy, then returns a
positive reweighting $H_0$ of edges of $G$ that is a
$(1\pm1/64)$-cut sparsifier of $G$. Since
$[63/64,65/64]\subseteq[24/25,26/25]$, the graph $H_0$
satisfies~\eqref{eq:coarse-accuracy}.

\paragraph{Failure probability.}
The construction calls Lemma~\ref{lem:query-recovery} $L+2$ times, once
for each $W_i$ and once for $w$, and Lemma~\ref{lem:query-forest} $L+2$
times, for stages $1,\ldots,L$, for $\widetilde Y$, and for
$\widetilde\Gamma$. Consider the first call that fails. Conditional on
the success of all earlier calls, its inputs satisfy the hypotheses of
its lemma and are independent of its own randomness, so it fails with
probability at most $\delta'$. Summing over the $2(L+2)$ calls, the
construction fails with probability at most $2(L+2)\delta'\le\delta_0$.
The parameter $\delta'$ of each copy already accounts for the
replacement of its random tapes.

\paragraph{Space and time.}
Both $U'=AU+m_0$ and $\delta'^{-1}$ are polynomial in $n$, as
Lemma~\ref{lem:query-recovery} requires, and there are $L+2=O(\log n)$
copies. By Lemma~\ref{lem:query-recovery},
each copy, including its seeds, uses $\widetilde O(n)$ bits, and each
recovered graph has $\widetilde O(n)$ edges. During recovery we keep
only the graph that supplies the current query cuts, the graph under
construction, and one query forest, so the workspace is also
$\widetilde O(n)$ bits.

Each update is applied to all copies, in
$\widetilde O(1)$ time per copy (Section~\ref{sec:proof-main}), so the
update time is $\widetilde O(1)$. After the
stream, each stage forms~\eqref{eq:regularizer-sketch} by rescaling the
$\widetilde O(n)$ counters of its copy and applying $n-1$ updates,
normalizes a graph with $\widetilde O(n)$ edges, checks connectivity,
and runs Lemmas~\ref{lem:query-forest} and~\ref{lem:query-recovery}.
Each of these steps takes $\widetilde O(n)$ time, so recovery takes
$\widetilde O(n)$ time over all $O(\log n)$ stages.
Appendix~\ref{app:precision} bounds the bit lengths of all numbers.

These bounds hold on every run: the limits on the number of selected
pairs, the counter widths, the loop bounds, and the time limit of
Lemma~\ref{lem:query-forest} are enforced on every run, and a run that
would exceed one of them aborts. Together with the approximation
guarantee and the failure probability, this proves
Theorem~\ref{thm:coarse}.

\subsection{Locally computable random tapes}\label{sec:tapes}

The analysis of Sections~\ref{sec:local}--\ref{sec:allcuts} assumes
that the survival bits and the evaluation signs are fully independent,
but storing them explicitly would take $\Omega(n^2)$ bits. We therefore
generate both sequences, which we call the survival tape and the
evaluation tape, from short seeds with the pseudorandom generator of
Kapralov et al.~\cite{KMM+20} (Proposition~\ref{prop:prg}). This
generator builds on the generators for space-bounded computation of
Nisan~\cite{Nisan92} and of Nisan and Zuckerman~\cite{NZ96}, and each of
its output bits can be computed in polylogarithmic time. We describe the construction for a single copy
of the three-table sketch, which sketches an integer vector
$x\in\{0,\ldots,U'\}^N$ with accuracy $\eps'$, failure parameter
$\delta'$, and sampling budget $\beta'$. This covers both the main
sketch and every copy in Section~\ref{sec:coarse}. Throughout, the query
cuts are fixed. They come from the coarse sketch and the query forest,
or, in Section~\ref{sec:coarse}, from earlier copies, so their
randomness is independent of both tapes. The same holds for the
four-wise seeds. Let $s_{\rm ctr}$ be the number of bits of the three
tables.

\paragraph{How the decoder uses the two tapes.}
On each tape, the bits are grouped into $N$ blocks, one per pair. Block
$e$ of the survival tape consists of $\xi_{e,1},\ldots,\xi_{e,J}$, and
block $e$ of the evaluation tape consists of all signs
$\sigma_e^{k,r,l}$. The decoder uses the two tapes differently. It never
reads a survival bit, since the survival bits affect the output only
through the counters of the three tables. In contrast, after selection
and the diagnostic, it reads the evaluation signs
$\sigma_e^{j_e,r_e,l}$, $l\in[R]$, of every selected edge. We therefore
replace the evaluation tape first, while the survival tape is still
truly random, and the survival tape second.

\paragraph{Success.}
Call a run of the decoder \emph{successful} if it returns a graph with
at most $\lceil8n\beta'\rceil$ edges that assigns positive weights only
to pairs of positive input weight and approximates every cut of $x$
within a factor of $1\pm\eps'$. Success captures the entire recovery
guarantee, so it suffices to show that each replacement decreases the
probability of success by at most $\delta'/20$. In particular, we do not
need the intermediate good events of
Sections~\ref{sec:local}--\ref{sec:allcuts} to keep their probabilities.

\paragraph{Algorithms for the analysis.}
Each replacement is justified by an algorithm that is used only in the
analysis. It reads one tape once, in order, has read-only access to all
other data, and finally decides whether the run is successful,
enumerating all cuts if necessary. Its working memory holds the tables,
the workspace of the decoder, and the signs stored in Step~1 below, so
it runs in space $S=O(s_{\rm ctr})$. We apply
Proposition~\ref{prop:prg} with $q=3$ and a constant $c$ such that
$S^{-c}\le\delta'/20$. Such a constant exists because $S\ge n$ and
$\delta'^{-1}\le n^{O(1)}$, and $q=3$ suffices because each tape has at
most $N(J+1)R^2\le S^3$ bits. The two replacements use independent
copies of the generator.

\paragraph{Step 1: the evaluation signs.}
Fix the query cuts, the survival tape, and all four-wise seeds. The
selection and diagnostic counters involve no evaluation signs, so the
candidates, the selected edges, their layers, and their rows are now
fixed. If more than $\lceil8n\beta'\rceil$ edges are selected, the run
is unsuccessful regardless of the evaluation signs. Otherwise, consider
the algorithm that visits the pairs $e=0,\ldots,N-1$ in order and reads
block $e$ of the evaluation tape. It adds the contribution of $e$, whose
weight $x_e$ it reads from its input, to every estimation counter, and if
$e$ is selected, it also stores the $R$ signs $\sigma_e^{j_e,r_e,l}$. By
linearity, its final counters equal those of the streaming algorithm.
They determine every estimate $\widehat w_e$, the test that all output
weights are positive, and hence the output and whether the run is
successful. The working memory consists of the three tables, at most
$\lceil8n\beta'\rceil R$ stored signs, and the workspace of the decoder
and of the final test. Proposition~\ref{prop:prg} therefore increases
the failure probability by at most $\delta'/20$ for every fixed choice
of the other randomness, and hence also after averaging over it.

\paragraph{Step 2: the survival bits.}
Now fix the query cuts, the four-wise seeds, and a seed of the
evaluation generator, so that the evaluation signs form a fixed string.
Consider the algorithm that visits the pairs in order, reads the bits
$\xi_{e,1},\ldots,\xi_{e,J}$ of block $e$ one at a time, and adds the
contribution of $e$ to all three tables at every layer $k$ with
$I_{e,k}=1$. Besides the three tables, its working memory holds only the
current pair and layer and the product of the bits read so far in the
current block. The final counters, together with the fixed seeds and
evaluation signs, determine everything the decoder does: selection and
candidate discovery use the selection counters, the diagnostic uses the
diagnostic counters, and the estimates use the estimation counters and
the fixed evaluation signs. Since the decoder never reads a survival
bit, the algorithm can finally run the decoder and decide success
within $O(s_{\rm ctr})$ bits of memory. An independent copy of the
generator therefore increases the failure probability by at most
another $\delta'/20$. Combining the two steps, the failure probability
with both generators exceeds its value with fully independent tapes by
at most $2\delta'/20$.

\paragraph{Computing the random bits.}
The sketch stores the two generator seeds, of
$O(S\operatorname{polylog}S)$ bits each, together with the four-wise
seeds. Whenever it processes an update to a pair $e$, or queries $e$
during recovery, it computes each bit of block $e$ that it needs from
the relevant seed, in $\operatorname{polylog}(n)$ time by
Proposition~\ref{prop:prg}. For the main sketch,
$s_{\rm ctr}=O(N_{\rm ctr}\log n)=\widetilde O(n/\eps^2)$, so both seeds
have $\widetilde O(n/\eps^2)$ bits. For a copy in
Section~\ref{sec:coarse}, they have $\widetilde O(n)$ bits. Both
replacement bounds hold for every fixed choice of accurate query cuts,
so they remain true conditional on the success of all earlier steps in
Section~\ref{sec:coarse}.

\subsection{Proof of Theorem~\ref{thm:main}}
\label{sec:proof-main}

If $n\le100$ or $N\le n/\eps^2$, we simply store the input exactly,
which meets all bounds of Theorem~\ref{thm:main}
(Appendix~\ref{app:precision}). Otherwise, we run the algorithm of
Section~\ref{sec:algorithm} with three changes: the decoder uses
candidate discovery (Section~\ref{sec:discovery}), the query cuts come
from the query forest of Lemma~\ref{lem:query-forest} with
$\delta_1=\delta/20$ (Section~\ref{sec:query-forest}), and the survival
and evaluation tapes are generated as in Section~\ref{sec:tapes}. As in
Section~\ref{sec:algorithm}, the coarse sketch is that of
Theorem~\ref{thm:coarse} with $\delta_0=\delta/20$. This theorem was
proved in Section~\ref{sec:coarse}.

\paragraph{Space.}
The three tables have $N_{\rm ctr}=\widetilde O(n/\eps^2)$ counters
(Section~\ref{sec:discovery}), each of $O(\log n)$ bits
(Appendix~\ref{app:precision}). By Theorem~\ref{thm:coarse}, the coarse
sketch uses $\widetilde O(n)$ bits, including its seeds and recovery
workspace. Normalizing $H_0$, computing the query forest
(Lemma~\ref{lem:query-forest}), and preprocessing it
(Section~\ref{sec:discovery}) need $\widetilde O(s_0+n)=\widetilde O(n)$
bits. Candidate discovery stores $N_{\rm id}=\widetilde O(n/\eps^2)$
decoded pairs (Section~\ref{sec:discovery}), and the decoder stores at
most $\lceil8n\beta\rceil$ selected pairs (Section~\ref{sec:algorithm}).
The generator seeds have $\widetilde O(n/\eps^2)$ bits
(Section~\ref{sec:tapes}), and the four-wise seeds have polylogarithmic
total size (Appendix~\ref{app:precision}). In total, the algorithm uses
$\widetilde O(n/\eps^2)$ bits.

\paragraph{Update time.}
Each update $(e,\Delta)$ is applied to the coarse sketch and to the
three tables at the two endpoints of $e$. Updating the coarse sketch
takes $\widetilde O(1)$ time by Theorem~\ref{thm:coarse}. For the
tables, we compute the $J$ survival bits of $e$ and, at each surviving
layer and each endpoint, update $R(d+1)$ selection, $64R^2$ diagnostic,
and $R^2$ estimation counters. This amounts to
$O(J(Rd+R^2))=O(\log^3n)$ counter updates, each of which needs one hash
value and one sign. Four-wise values and generator bits take
polylogarithmic time to compute (Appendix~\ref{app:precision} and
Proposition~\ref{prop:prg}), so the worst-case update time is
$\widetilde O(1)$.

\paragraph{Recovery time.}
Computing $H_0$ takes $\widetilde O(n)$ time by
Theorem~\ref{thm:coarse}, normalizing it takes $O(s_0\log s_0+n)$ time
(Appendix~\ref{sec:normalize}), and computing the query forest takes
$\widetilde O(s_0+n)=\widetilde O(n)$ time by
Lemma~\ref{lem:query-forest}. Recovery from the tables then takes
$\widetilde O(n/\eps^2)$ time (Section~\ref{sec:discovery}). These
bounds hold on every run, so the recovery time is
$\widetilde O(n/\eps^2)$ in the worst case.

\paragraph{Correctness.}
The coarse sketch fails with probability at most $\delta/20$ by
Theorem~\ref{thm:coarse}, and so does the algorithm of
Lemma~\ref{lem:query-forest}. If neither fails, then the query cuts are
accurate, and they are independent of the three tables in any case
(Section~\ref{sec:query-forest}). Fix accurate query cuts, and suppose
first that the tapes are fully independent. By
Proposition~\ref{prop:correctness}, outside eight events of probability
at most $\delta/20$ each, pair enumeration returns a positive
reweighting $H$ of a set $E_\star\subseteq E$ of at most
$8n\beta=O(n\eps^{-2}\log^2n)$ edges, and $H$ is a $(1\pm\eps)$-cut
sparsifier of $G$. In particular, the load bound and selection success
hold outside these events. Proposition~\ref{prop:discovery} then shows
that, except with probability $\delta/20$, candidate discovery returns
the same output. Hence a run is successful in the sense of
Section~\ref{sec:tapes} except with probability $9\delta/20$.
Generating the tapes as in Section~\ref{sec:tapes} increases this
failure probability by at most $2\delta/20$, to at most $11\delta/20$.
The total failure probability is therefore at most
$\delta/20+\delta/20+11\delta/20=13\delta/20<\delta$. Together with the
bounds on space, update time, and recovery time, this proves
Theorem~\ref{thm:main}.

\section*{Acknowledgements}
The authors began studying this problem in 2025 and developed an initial
solution, in which an issue was later identified. Subsequently,
Stellar-Colosseum~\cite{colosseum}, a fully automated agentic system
built on the Gemini~4 base model, proposed a new end-to-end proof for an
algorithm with near-optimal space complexity. Following further
interaction, the Gemini model also proposed a way to achieve near-linear
running time. The authors verified the proof, edited it for clarity of presentation, and take
responsibility for the final version. Other
models were subsequently used to edit the text. The authors thank
Oswaldo Ramirez for helpful discussions during the early stages of this
work.

\bibliographystyle{alpha}
\bibliography{reference}

\appendix

\section{Implementation details}\label{app:precision}

This appendix gives the implementation details omitted from the main
text: the small cases, the bit lengths of all numbers, the normalization
of the graphs used for query cuts, and the four-wise independent hash
functions and signs.

\subsection{Small cases and bit lengths}

We first round $\eps$ and $\delta$ down to powers of two. This changes
each of them by at most a factor of two and affects only constants.

If $n\le100$ or $N\le n/\eps^2$, we store all $N$ coordinates exactly in
an array indexed by $\id$ and output $G$ by scanning the array. This
uses $O(N\log(MU+2))$ bits, $O(1)$ time per update, and $O(N)$ recovery
time. For $n\le100$, the space is $O(\log(MU+2))$ bits and both times
are $O(1)$. For $n>100$, we have $N\le\min\{n^2,n/\eps^2\}$, so the space is
$O(\min\{n^2,n/\eps^2\}\log n)$ bits, the recovery time is
$O(\min\{n^2,n/\eps^2\})$, and the output has at most $n/\eps^2$ edges.

In the remaining case, $n/\eps^2<N<n^2$, so $\eps^{-1}<\sqrt n$.
Together with $M,U,\delta^{-1}\le n^{O(1)}$, this implies that every
parameter has an $O(\log n)$-bit representation. Moreover, the bounds
$\widetilde O(n/\eps^2)$ on space and recovery time are also
$\widetilde O(\min\{n^2,n/\eps^2\})$. The analysis uses $\ell$ only
through $\ell\ge\ln(64n^2J/\delta)$ and $\ell=O(\log n)$, so the
implementation may take $\ell=\lceil\log_2(64n^2J/\delta)\rceil$, which
can be computed exactly.

Every coefficient of a three-table sketch lies in $\{-1,0,1\}$, so all
counters are integers. Recovery computes medians, squares of difference
queries, averages of $R$ counters, and rescalings by $2^k$, and compares
these values with thresholds determined by $\kappa_e$. Every $\kappa_e$
is the weight of a cut in a normalized graph
(Section~\ref{sec:normalize}) or, at the first stage of
Section~\ref{sec:coarse}, the integer $m_0$. All of these numbers are
therefore rationals with $O(\log n)$-bit numerators and denominators,
and each operation on them takes $O(1)$ time.

\subsection{Counter widths}

During the stream, each update changes every counter by at most $U$, so
every vertex counter of every copy has magnitude at most $MU$. In
Section~\ref{sec:coarse}, the vertex counters formed
in~\eqref{eq:regularizer-sketch} have magnitude at most
$AMU+m_0(n-1)$. The algorithms of Section~\ref{sec:tapes} build the
counters coordinate by coordinate, so every partial sum they produce
has magnitude at most $NU'$. Postorder aggregation
(Section~\ref{sec:discovery}) sums at most $n$ vertex counters. Hence
every counter has magnitude at most $nM_{\max}$, where
$M_{\max}=\max\{AMU+m_0n,\,NU'\}\le n^{O(1)}$, and $O(\log n)$ bits per
counter suffice.

\subsection{Normalizing the graphs used for query cuts}
\label{sec:normalize}

Query cuts are computed from recovered graphs. We normalize these graphs
so that the computation is exact and fast, even if the recovery that
produced a graph has failed. For an integer bound $V'$ and a graph $Y$,
let $\mathcal R_{V'}(Y)$ be obtained from $Y$ by merging parallel
entries, replacing each weight $Y_e$ by
\[
 2^{-\mu}\Bigl\lfloor
 2^\mu\min\{\max\{Y_e,0\},2nV'\}\Bigr\rfloor,
 \qquad \mu=\lceil\log_2(100n^2)\rceil,
\]
and discarding entries that become zero. For a graph with $s$ entries,
this takes $O(s\log s+n)$ time: sort the entries by identifier, merge
parallel entries, and round. The clipping to $[0,2nV']$ bounds every
weight on every run, and the next lemma shows that it has no effect when
$Y$ is a good approximation.

\begin{lemma}[Rounding a cut approximation]\label{lem:rounding}
Let $x\in\{0,1,\ldots,V'\}^N$, and let $Y$ be a nonnegative graph that
approximates every cut of $x$ within $[24/25,26/25]$. Then
$\mathcal R_{V'}(Y)$ approximates every cut of $x$ within $[.95,1.04]$
and has the same connected components as $x$. Its weights are positive
multiples of $2^{-\mu}$ and at most $2nV'$.
\end{lemma}
\begin{proof}
Write $Y'=\mathcal R_{V'}(Y)$ and $x(S)=x(S,V\setminus S)$, and
similarly for $Y$ and $Y'$.

\emph{Clipping has no effect.} For a pair $e=uv$,
$Y_e\le Y(\{u\})\le\frac{26}{25}x(\{u\})\le\frac{26}{25}(n-1)V'<2nV'$.

\emph{Rounding loses little.} Rounding down to a multiple of
$2^{-\mu}\le1/(100n^2)$ decreases each weight by less than
$1/(100n^2)$. A cut is crossed by fewer than $n^2/2$ pairs, so
$Y(S)-1/200<Y'(S)\le Y(S)$ for every $S$.

\emph{Cut approximation.} If $x(S)=0$, then
$Y(S)\le\frac{26}{25}x(S)=0$, so $Y'(S)=0$. Otherwise $x(S)\ge1$
because $x$ is integral, and
\[
 .95\,x(S)\le\frac{24}{25}x(S)-\frac{x(S)}{200}<Y'(S)\le Y(S)
 \le\frac{26}{25}x(S)=1.04\,x(S).
\]

\emph{Components and weights.} A set $S$ is a union of components of
$x$ exactly when $x(S)=0$, which by the previous step happens exactly
when $Y'(S)=0$. Hence $Y'$ and $x$ have the same components. The
weights of $Y'$ have the stated form by construction.
\end{proof}

We apply this normalization with $V'=U$ to the output of
Theorem~\ref{thm:coarse} (Section~\ref{sec:query-forest}), and
Section~\ref{sec:coarse} applies it with the weight bound of each copy.

\subsection{Four-wise independent hash functions and signs}

Let $m$ be the smallest integer of the form $2\cdot3^a$ with
$2^m\ge\max\{N,B_S\}$. Then $m=O(\log n)$. The polynomial
$z^m+z^{m/2}+1$ is the cyclotomic polynomial $\Phi_{3^{a+1}}$, and it is
irreducible over $\mathbb F_2$ because $2$ is a primitive root modulo
$3^{a+1}$ (see, e.g.,~\cite{LN97}). We represent $\mathbb F_{2^m}$ as
$\mathbb F_2[z]/(z^m+z^{m/2}+1)$, so a field element occupies $O(1)$
words and each field operation takes $O(\log n)$ time. We identify each
pair $e$ with the field element whose binary representation is
$\id(e)$. A uniformly random polynomial of degree at most three over
$\mathbb F_{2^m}$, evaluated at the pairs, gives values that are
four-wise independent and uniform on $\mathbb F_{2^m}$. Every table
width is a power of two at most $2^m$. For a table of width $B$,
$\log_2B$ fixed bits of the value therefore give a four-wise independent
hash into $[B]$ with uniform marginals, and one fixed bit gives four-wise independent unbiased signs.
Each function has a seed of $4m=O(\log n)$ bits and takes $O(1)$ field
operations to evaluate. A copy of the sketch uses $O(JR^2)$ hash
functions and sign families, dominated by the $64R$ diagnostic sign
families of each layer and row, so its four-wise seeds take
$O(JR^2\log n)$ bits.

\end{document}